\documentclass[12pt, letterpaper]{elsarticle}

\makeatletter
\def\ps@pprintTitle{%
 \let\@oddhead\@empty
 \let\@evenhead\@empty
 \let\@oddfoot\@empty
 \let\@evenfoot\@empty}
\makeatother

\usepackage{graphicx}
\usepackage{bm}
\usepackage{diffcoeff} 
\usepackage{enumitem}
\usepackage[margin=1.2in]{geometry}
\usepackage{caption}
\usepackage{subcaption}
\usepackage[utf8]{inputenc}
\usepackage[T1]{fontenc}
\usepackage[english]{babel}
\usepackage{textcomp}
\usepackage{latexsym, amsmath, amssymb}
\usepackage{booktabs, multirow}
\usepackage[version=4]{mhchem}

\usepackage{mathtools}
\usepackage{exscale, relsize}
\usepackage[retainorgcmds]{IEEEtrantools}
\usepackage{algpseudocode}
\usepackage{siunitx}
\DeclareSIUnit\atm{atm}

\usepackage[hyphens]{url}
\usepackage[breaklinks=true, linkcolor=blue, citecolor=blue, colorlinks=true]{hyperref}

\allowdisplaybreaks

\date{}

\begin{document}

\begin{frontmatter}

\title{Uncertainty quantification of reacting fluids interacting with porous media using a hybrid physics-based and data-driven approach}

\author[osu]{Diba Behnoudfar}
\author[osu]{Kyle E.~Niemeyer}
\address[osu]{School of Mechanical, Industrial, and Manufacturing Engineering, Oregon State University, Corvallis, Oregon, USA}

\begin{abstract}
Accurately simulating coupled physical processes under uncertainty is essential for reliable modeling and design in performance-critical applications such as combustion systems.
Ablative heat shield design, as a specific example of this class, involves modeling multi-physics interactions between reacting flows and a porous material.
Repeatedly evaluating these models to quantify parametric uncertainties would be prohibitively computationally expensive. 
In this work, we combine physics-based modeling using a single-domain approach with data-driven reduced-order modeling to quantify uncertainty via the operator inference method.
The detailed physics-based simulations reproduce the measured surface temperature of an object exposed to high-enthalpy flow in a plasma wind tunnel experiment within 5\%.
We further use the model to demonstrate the effect of complex flow situations on the dynamic interactions between the porous heat shield material and the surrounding gas.
The parametric reduced-order model, built on physics-based simulation data, successfully captures variations in quantities of interest resulting from changes in the permeability and heat transfer coefficient of the porous material in two separate studies: solid-fuel combustion with emission of buoyant reacting plumes in quiescent air, and ablation in a wind tunnel. 
\end{abstract}


\end{frontmatter}

\section{Introduction}
Most physical processes can be represented by a set of governing equations in the form of parametrized partial differential equations (PDEs), where a set of design parameters control the model predictions, such as physical properties, the boundary conditions, or the geometry of the computational domain.
For applications such as uncertainty quantification or design optimization, a large number of parameters must be evaluated.
Detailed simulations of large systems, particularly those involving coupled processes such as momentum and energy transfer, are typically resource-intensive, demanding substantial memory and significant computational time. 

In the context of reacting flows, combustion simulations are affected by several sources of uncertainty, which degrade the reliability of model predictions \cite{mueller2013chemical}. 
For example, in the design of ablative thermal protection systems, determining the appropriate size of the heat shield is a crucial part of developing an atmospheric entry vehicle.
During entry, the vehicle experiences intense aerodynamic heating due to the conversion of significant kinetic energy into heat \cite{mansour2024flow}. 
An ablative thermal shield protects the vehicle through thermo-chemo-mechanical degradation processes.
Initial design and performance evaluations rely on physics-based models to estimate the heat loads.
These models can involve multiple uncertain parameters, including the transport properties of the porous heat shield material, which are generally difficult to estimate due to its complex geometry.
Consequently, evaluating models' predictive capabilities becomes impossible without quantifying the uncertainty in predictions. 

Traditional methods for quantifying uncertainty mainly revolve around Monte Carlo simulation \cite{caflisch1998monte} by brute-force sampling of the input parameter distributions and propagating them through the problem.
This method of forward uncertainty quantification is straightforward but expensive, and it has a slow rate of convergence, meaning that a large sample size is required to ensure good accuracy \cite{kolla2022comprehensive}.
Given the computational expense of simulations, uncertainty quantification has relied heavily on surrogate models to approximate the original model \cite{kolla2022comprehensive}.
Among the most popular surrogate approaches in uncertainty quantification are polynomial chaos expansions \cite{ghanem2003stochastic,xiu2002wiener}, stochastic collocation \cite{xiu2005high}, and Gaussian processes \cite{williams2006gaussian}. 

In combustion research, many studies have focused on propagating uncertainty in chemical kinetics through gas-phase models, often using polynomial chaos expansion (PCE) to build surrogate models—primarily with non-intrusive methods \cite{sheen2009spectral, reagana2003uncertainty}, though some used intrusive stochastic Galerkin approaches \cite{silva2020uncertainty, reagan2004spectral}.
The intrusive methods directly modify the governing equations of the model to incorporate uncertainty. This often requires access to and changes in the internal structure of the simulation code.
Non-intrusive methods, on the other hand, treat the existing simulation model as a ``black box'' and do not require changes to the original code.
In early studies, chemical kinetic uncertainty quantification was mostly applied to simple systems like homogeneous reactors and one-dimensional laminar flames.
Mueller et al.~\cite{mueller2013chemical} advanced this by extending the analysis to turbulent non-premixed combustion, introducing a physics-informed dimension reduction method that mapped kinetic uncertainty onto flamelets and reduced the problem to a single parameter by assuming a correlation between temperature and scalar dissipation rate.
Others later combined projection-based dimensionality reduction techniques such as principal component analysis (PCA) \cite{vajda1985principal} and active subspace with the surrogate modeling to mitigate the high dimensionality problem \cite{koenig2023kinetic, vohra2019active}.
In recent years, artificial neural networks have also been extensively used to construct surrogate models \cite{wang2020facilitating, koenig2023kinetic, cai2021physics}. 
Apart from reaction kinetics parameters, uncertainty propagation due to other input parameters such as boundary conditions and scalar mixing has also been investigated \cite{enderle2020non, amaduzzi2023impact}.
The above-mentioned methods such as PCE and Gaussian processes struggle with transient problems as they do not inherently adapt to evolving dynamics in time-dependent systems.
As time progresses, the number of terms required in the PCE increases rapidly to maintain accuracy, leading to computational challenges \cite{branicki2013fundamental}.
With Gaussian processes, specialized kernels or approximations are required.
Furthermore, the inversion of the covariance matrix, which is central to Gaussian processes inference, has a computational complexity of $\mathcal{O}(n^3)$, where $n$ is the number of data points. 
This cubic scaling limits their applicability to small- to medium-sized datasets \cite{williams2006gaussian}. 
On the other hand, surrogate models based on neural networks---including physics-informed neural networks---do not explicitly incorporate the known polynomial structure of the governing equations and can struggle with strongly coupled, high-dimensional multiphysics problems \cite{cai2021physics}. 
Furthermore, neural operator architectures typically require large training datasets and substantial computational resources for training, which can be prohibitive for the class of coupled reacting flow--porous media problems considered in this work.

Operator inference \cite{peherstorferhar2016data,kramer2019nonlinear,qianReducedOperatorInference2022,kramerLearningNonlinearReduced2024} is a non-intrusive approach for reducing models of systems that exhibit polynomial nonlinearities.
It is designed to learn the differential equations (or operators) that govern the evolution of a physical system---but in a reduced form.
Similar to the aforementioned projection-based methods, the operator inference technique uses singular value decomposition to define a low-dimensional subspace (or coordinates) from the full-order model (FOM) calculations.
Compared to methods such as PCE, operator inference can handle large-scale, high-dimensional systems more effectively, especially when those systems have underlying low-dimensional dynamics \cite{swischuk2020learning}.
Polynomial chaos expansions become computationally prohibitive as the number of uncertain parameters increases due to the curse of dimensionality.
In addition, operator inference is well-suited to capturing transient dynamics and time-evolving behavior of systems as it directly learns the operators governing system evolution.
The structure of the reduced-order model (ROM) based on operator inference is guided by the known governing equations, and the ROM operators are determined by minimizing a data-driven residual within the reduced state space.
The method has been used to build ROMs for control of combustion systems \cite{swischuk2020learning}, among other applications
in fluid dynamics, aeroelasticity, mechanical\slash multi-body systems, atmospheric science, and geosciences~\cite{peherstorferhar2016data, zastrow2023data,filanova2023operator,rocha2023reduced, gkimisis2025non}.

Phenomena such as ablation under high-enthalpy flow conditions or fires in built or natural environments occur across multiple regions with strong coupling between the processes near the interface.
Capturing this dynamic interaction involving surface chemistry, radiation, and fluid flow is essential for predictive modeling tools, especially in estimating heat fluxes in heat shield applications.
Since new low-density ablative materials are highly porous \cite{mansour2024flow}, the flow at the interface between the surrounding gas and the porous material can exhibit a variety of behaviors, depending on the flow regime; for example, 
under laminar regimes, the porous interface coupling and through-flow can lead to the formation of flow instabilities, whereas in turbulent flows, eddies may penetrate into the porous domain, promoting enhanced interfacial transport.
Despite advancements in this field, the interaction between porous heat shields
and the surrounding complex flow remains poorly understood \cite{mansour2024flow}.
In a previous study \cite{behnoudfar2025single}, we demonstrated that a single-domain modeling approach is effective for capturing these interfacial processes in the problem of solid-fuel combustion and near-field flow dynamics. In this approach, transport phenomena are captured through spatially averaged equations, enabling a uniform representation across the entire domain. 

Applying parametric operator inference to this class of problems is non-trivial for several reasons.
The governing equations contain non-polynomial reactive source terms that must be approximated, and the coupled multi-physics system requires careful selection of lifted variables (auxiliary variables introduced to convert non-polynomial terms into a polynomial state-space structure) to achieve a polynomial structure.
The affine parametric dependence (where the influence of parameters can be factored out of the system operators) must be identified from the physical structure of the equations.
Furthermore, as we will show, the construction of parametric ROMs for this system reveals unexpected sensitivities to the number and placement of training parameter samples.

In this paper, we apply the single-domain approach to the ablation problem (serving as the full-order model) and use the resulting simulation data---along with data from a previous study \cite{behnoudfar2025single}---to develop a parametric operator inference framework for quantifying uncertainty in solid fuel combustion simulations. 
We first introduce the methods underlying the detailed physics-based simulations, then the reduced-order modeling framework. 
Next, we apply the method to two separate studies: solid fuel combustion in quiescent air and ablation in a wind tunnel. 
We select permeability and heat transfer coefficient in the porous region as uncertain parameters and showcase the model's applicability to uncertainty quantification for this class of problems. 

\section{Methodology}

\subsection{Governing equations with a single-domain approach \label{subsec:gequation}}
The processes in the porous region and the surrounding fluid are modeled using the single-domain approach, which uses a single set of transport equations for the fluid and solid in the entire domain, as described in more detail by Behnoudfar and Niemeyer~\cite{behnoudfar2025single}.
For the fluid and solid phases, the model solves the following volume-averaged conservation equations:
\begin{align}
\diffp{}{t} \left(\phi \langle \rho \rangle^f\right) + \nabla \cdot \left(\phi \langle \rho \rangle^f \langle \textbf{u} \rangle\right) &= (1-\phi)\sum_{j=1}^{N_g} \langle \dot \omega^s_{g,j} \rangle^s \;, \label{eq:1} \\
\diffp{}{t} \left(\phi \langle \rho \rangle^f \langle Y_{g,j} \rangle \right) + \nabla\cdot \left(\phi \langle \rho \rangle^f \langle Y_{g,j} \rangle \langle \textbf{u} \rangle\right) &= \nabla\cdot(\phi \langle \rho \rangle^f D \nabla \langle Y_{g,j} \rangle) -\> \nabla\cdot\left(\phi \langle \rho \rangle^f \langle \hat{Y}_{g,j}  \hat{\textbf{u}} \rangle \right) \nonumber\\
 &+ \phi \langle \dot\omega_{g,j} \rangle^f + \left(1-\phi\right) \langle \dot\omega_{g,j}^s \rangle^s \;,  \label{eq:2} \\
\diffp{}{t} \left(\phi \langle \rho \rangle^f \langle \textbf{u} \rangle\right) + \nabla\cdot\left(\phi \langle \rho \rangle^f \langle \textbf{u} \rangle \langle \textbf{u} \rangle \right) + \nabla \langle p \rangle &= \nabla\cdot \langle \boldsymbol{\tau}\rangle - \nabla\cdot\left(\phi \langle \rho \rangle^f \langle \hat{\textbf{u}}  \hat{\textbf{u}} \rangle \right) +\> \phi \langle \rho \rangle^f \textbf{g} + \textbf{f} \;, \label{eq:3} \\
\diffp{}{t} \left(\phi \langle \rho \rangle^f \langle h \rangle \right) + \nabla\cdot(\phi \langle \rho \rangle^f \langle h \rangle \langle \textbf{u} \rangle) &= \nabla\cdot( \lambda \nabla \langle T \rangle) + \dot Q -\> \nabla\cdot(\phi \langle \rho \rangle^f \langle \hat{h}  \hat{\textbf{u}} \rangle)\nonumber\\
 &-\> h_{fs} A_{fs} \left(\langle T \rangle - \langle T^s \rangle\right) + \diffp{\langle p \rangle}{t} \nonumber \\
&+\> (1-\phi) \sum_{j=1}^{N_g} \langle h_{g,j} \rangle \langle \dot\omega_{g,j}^s \rangle^s + \langle S^{f,\> \text{rad}} \rangle\;, \label{eq:4} \\
\diffp{}{t} \left((1-\phi) \langle \rho \rangle^s\right) &= (1-\phi)\sum_{i=1}^{N_s} \langle \dot \omega^s_{s,i} \rangle^s \;, \label{eq:5} \\
\diffp{}{t} \left((1-\phi) \langle \rho \rangle^s \langle Y_{s,i} \rangle \right) &= (1-\phi) \langle \dot \omega^s_{s,i} \rangle^s \;, \label{eq:6} \\
\diffp{}{t} \left((1-\phi) \langle \rho \rangle^s \langle h^s \rangle \right) &= \nabla\cdot( \lambda^s \Bar{\Bar{A}}\nabla \langle T^s \rangle)
-\> (1-\phi) \sum_{i=1}^{N_s} \Delta h^\circ_{s,i} \langle \dot\omega_{s,i}^s \rangle^s \nonumber\\
&+\> h_{fs} A_{fs} \left(\langle T \rangle - \langle T^s \rangle\right)  \nonumber\\
&-\> (1-\phi) \sum_{j=1}^{N_g} \langle h_{g,j} \rangle \langle \dot\omega_{g,j}^s \rangle^s 
+ \langle S^{s,\> \text{rad}} \rangle \label{eq:7}
\end{align}
where $\textbf{u}$ is the fluid velocity vector, $t$ is time, $T$ is fluid temperature, $T^s$ is solid temperature, $\phi$ is the porosity, $Y_{g,j}$ is mass fraction of gas species $j$, $\dot\omega_{g,j}$ is volumetric mass change rate of gas species $j$ resulting from homogeneous reactions, $\dot\omega_{g,j}^s$ is volumetric mass change rate of gas species $j$ resulting from heterogeneous reactions, $p$ is fluid pressure, 
$\rho$ is density, $D$ is the fluid diffusion coefficient, 
$h$ is the fluid's mixture-averaged enthalpy, $h_{g,j}$ is the enthalpy of gas species $j$, $\lambda$ is fluid thermal conductivity, $\dot Q = -\phi \sum_{j=1}^{N_g}\langle \dot\omega_{g,j} \rangle^f \Delta h^\circ_{g,j}$ is the heat release rate due to gas-phase reactions, $N_g$ is the number of gas species, $\Delta h^\circ_{g,j}$ is the heat of formation of individual gas species $j$, $\textbf{g}$ is gravitational acceleration, and $S^{f,\> \text{rad}}$ is the heat exchanged through radiation for the gas phase, evaluated using a participating media approach.
The symbols $\langle \cdot \rangle$, $\langle \cdot \rangle^f$ and $\langle \cdot \rangle^s$ denote volume averaging operators defined as $\langle \cdot \rangle = 1/\Delta V\int_{\Delta V} (\cdot)\,dV$; 
for $\langle \cdot \rangle^f$ and $\langle \cdot \rangle^s$, the total averaging volume ($\Delta V$) is replaced with $\Delta V^f$ or $\Delta V^s$, which are the volumes of the fluid and solid, respectively.
$\hat{\textbf{u}}$, $\hat{h}$, and $\hat{Y}_{g,j}$ are the fluctuating components of velocity, enthalpy, and mass fraction, respectively, due to averaging defined as  $\hat{\textbf{u}}= \textbf{u} - \langle\textbf{u}\rangle$, $\hat{h}= h - \langle h \rangle$ and  $\hat{Y}_{g,j}= Y_{g,j} - \langle Y_{g,j} \rangle$.
$\boldsymbol{\tau}$ is the stress tensor:
\begin{equation}
\langle \boldsymbol{\tau}\rangle = \mu^{*}\left(\nabla \langle\textbf{u}\rangle+{\nabla\langle\textbf{u}\rangle}^T\right)-\frac{2}{3}\mu^{*}\left(\nabla \cdot \langle\textbf{u}\rangle\right)\textbf{I}\;,
\end{equation}
where $\mu^*$ is the effective viscosity \cite{le2006}; in this work, we assume $\mu^* = \mu$ (i.e., the fluid viscosity) for simplicity.
$\textbf{f}$ represents the drag force that the solid exerts on the fluid phase,
$ \textbf{f}=\frac{1}{\Delta V}\int_{\Delta A}(-p\mathbf{I}+\boldsymbol{\tau})\cdot\textbf{n}\,dA \;,$ where $\Delta A$ is the interfacial area between the fluid and solid phases in $\Delta V$ and $\textbf{n}$ is the normal vector.
This term is commonly modeled using the correlation form $\textbf{f} = - \mu \textbf{K}^{-1}[\textbf{I}+\text{Re}_p \textbf{F}_0]\cdot \langle\textbf{u}\rangle$, where $\mathbf{K}$ is Darcy permeability tensor (which is a function of porosity),  $\text{Re}_p = \frac{\langle \rho \rangle^f  ||\langle\textbf{u}\rangle|| d_p}{\mu}$ is pore Reynolds number where $d_p$ is the pore diameter and $\textbf{F}_0$ is the Forchheimer coefficient tensor.
The other non-closed term, $\nabla\cdot\left(\phi \langle \rho \rangle^f \langle \hat{\textbf{u}}  \hat{\textbf{u}} \rangle \right)$, appears because velocity fluctuates inside the control volume and thus differs from its averaged value. 
The quantity $\boldsymbol{\tau}_{\textrm{SGS}}=-\left(\phi \langle \rho \rangle^f \langle \hat{\textbf{u}}  \hat{\textbf{u}} \rangle \right)$  is a part of the sub-filter scale stress (or sub-grid scale, SGS) and, according to large-eddy simulation techniques, is commonly modeled as 
$-\left(\phi \langle \rho \rangle^f \langle \hat{\textbf{u}}  \hat{\textbf{u}} \rangle \right)= \mu_t\left(\nabla\langle\textbf{u}\rangle + \left(\nabla\langle\textbf{u}\rangle\right)^T\right)$  where $\mu_t$ is turbulence eddy viscosity.
The term $h_{fs} A_{fs} (\langle T \rangle - \langle T^s \rangle)$ approximates spatially averaged convective heat transfer between the fluid and solid inside the porous region, where $h_{fs}$ is the heat transfer coefficient and $A_{fs}$ is the interfacial surface area between fluid and solid. 
$N_s$ is the number of solid species, $\dot\omega_{s,i}^s$ is volumetric mass rate of change for solid species $i$ resulting from heterogeneous reactions, $Y_{s,i}$ is the mass fraction of solid species $i$, $h^s$ is the solid's mixture-averaged enthalpy, $\Delta h^{\circ}_{s,j}$ is the heat of formation of solid species $j$, $\lambda^s$ is the solid's mixture-averaged thermal conductivity, $\Bar{\Bar{A}}$ is the anisotropy tensor of the solid matrix, and $S^{s,\> \text{rad}}$ is the heat exchanged through radiation for the solid phase.
Using the unity Schmidt and Lewis number assumptions, diffusion coefficient and thermal conductivity are obtained from the viscosity evaluated using Sutherland's model \cite{sutherland1893}.

The advective terms in the momentum equation are discretized using a blend of central differencing (75\%) and linear upwind (25\%) schemes to interpolate the velocity at cell faces.
The advective terms in the energy and species conservation equations use a central differencing scheme that limits towards upwind in regions of rapidly changing gradients according to the procedures in the total variation diminishing (TVD) schemes \cite{harten1997high}. The diffusive terms are discretized using standard central differencing along with linear interpolation for the diffusivity.
The solver in this work is based on the \texttt{reactingFoam} solver of OpenFOAM~\cite{openfoam}, which uses a sequential splitting method in which the reactive terms are integrated separately. 
We use an extrapolation-based stiff ordinary differential equation (ODE) solver (SEULEX)~\cite{hairer1991ii} to integrate the solid-phase reactive terms and a L-stable, stiff embedded Rosenbrock ODE solver of order (2)3~\cite{sandu1997benchmarking} to integrate the gas-phase reactive terms. 
Further details of the solution algorithm are discussed by Behnoudfar and Niemeyer~\cite{behnoudfar2025single}. 

\subsection{Operator inference for learning reduced-order models}
The system of discretized partial differential equations described by Eqs.~\eqref{eq:1}--\eqref{eq:7} can be written as
\begin{equation}
    \diff{\textbf{q}}{t} = \textbf{F}\left(\textbf{q};\bm{\eta} \right),
    \label{eq:states}
\end{equation}
where $\textbf{q} \in \mathbb{R}^{dn_{xy}} $ is the vector of concatenated $d$ state variables at $n_{xy}$ grid points
\[
\mathbf{ \textbf{q}}( \bm{\eta}) = \begin{bmatrix} 
\mathbf{s_1} \\ 
\vdots \\ 
\mathbf{s_{n_{xy}}}
\end{bmatrix},
\]
\begin{equation*}
  \textbf{s}_i = \left[\langle u_x \rangle \;\;\langle u_y \rangle \;\; \langle T \rangle \;\;\langle T_s \rangle \;\; \phi\langle \rho \rangle^f \;\; \langle Y_{g,1} \rangle \dots  \langle Y_{g,N_g}\rangle \;\; \langle Y_{s,1} \rangle \dots  \langle Y_{s,N_s}\rangle\right]^T \bigg|_{i,\bm{\eta}} \in \mathbb{R}^d,
\end{equation*}
$\textbf{F}: \mathbb{R}^{dn_{xy}} \times \mathbb{R}^{m} \rightarrow \mathbb{R}^{dn_{xy}}$ represents the spatial discretization of the system of equations and maps $\textbf{q}$ and $\bm{\eta}$ to time derivatives of $\textbf{q}$ and $\bm{\eta} \in \mathbb{R}^m$ is the vector of $m$ parameters on which the system depends. 
Note that the left-hand sides of Eqs.~\eqref{eq:1}--\eqref{eq:7} are given as functions of $\phi \langle \rho \rangle^f \langle Y \rangle$, $\phi \langle \rho \rangle^f \langle u_x \rangle$, $\phi \langle \rho \rangle^f \langle h \rangle$, etc., but the equations can be rewritten in terms of the above state variables ($\textbf{s}_i$) to allow for more flexibility and easier computations \cite{swischuk2020learning}. In the following applications, we assume a two-dimensional model, though this can be extended to three dimensions.

We consider the setting in which $\textbf{F}$ has a polynomial structure; many partial differential equations exhibit this structure or can be written in this form through a change of variables \cite{mcquarrie2023nonintrusive}. Thus, the system represented by Eq.~\eqref{eq:states} can be described in polynomial form as
\begin{equation}
    \diff{\textbf{q}}{t} = \textbf{A}(\bm{\eta})\textbf{q} + \textbf{B}(\bm{\eta})(\textbf{q} \otimes \textbf{q}) +  \textbf{C}(\bm{\eta})(\textbf{q} \otimes \textbf{q} \otimes \textbf{q}) + \textbf{c}(\bm{\eta}) + \dots \;,
    \label{eq:poly}
\end{equation}
where $\textbf{A}(\bm{\eta})\textbf{q}$ represents the terms in $\textbf{F}(\cdot)$ that are linear in $\textbf{q}$, 
$\textbf{B}(\bm{\eta})(\textbf{q} \otimes \textbf{q})$ represents the terms that are quadratic in $\textbf{q}$ (for example, $\nabla \cdot \left(\phi \langle \rho \rangle^f \langle \textbf{u} \rangle\right)$ in Eq.~\eqref{eq:1} is quadratic in states $\phi \langle \rho \rangle^f$ and $\langle \textbf{u} \rangle$ ), 
$\textbf{C}(\bm{\eta})(\textbf{q} \otimes \textbf{q} \otimes \textbf{q})$ represents the terms cubic in $\textbf{q}$, and 
$\textbf{c}(\bm{\eta})$ represents the constant terms, with $\textbf{A} \in \mathbb{R}^{dn_{xy} \times dn_{xy}}$, $\textbf{B} \in \mathbb{R}^{dn_{xy} \times (dn_{xy})^2}$, $\textbf{C} \in \mathbb{R}^{dn_{xy} \times (dn_{xy})^3}$ and $\textbf{c} \in \mathbb{R}^{dn_{xy}}$ as the respective operators.
In addition $\otimes$ is the Kronecker product \cite{guo2022bayesian}.

For the terms with a non-polynomial structure, such as the reactive source terms, new auxiliary variables can be introduced through the process of ``lifting the equation'' \cite{kramer2019nonlinear}, so that the governing equations become polynomial in the lifted states. A complete lifting that converts all equations to a polynomial form is possible but would require the introduction of a large number of auxiliary variables. As shown by Swischuk et al. \cite{swischuk2020learning}, many terms in this system of equations take a polynomial form if the following variables are used:
\begin{equation*}
    \textbf{s}_L = \left[\langle u_x \rangle \;\;\langle u_y \rangle \;\; \langle T \rangle \;\;\langle T_s \rangle \;\; \frac{1}{\phi\langle \rho \rangle^f} \;\; \langle c_{g,1} \rangle \dots  \langle c_{g,N_g}\rangle \;\; \langle Y_{s,1} \rangle \dots  \langle Y_{s,N_s}\rangle\right]^T,
\end{equation*}
where $c_{g,j} = \frac{\phi\langle \rho \rangle^f \langle Y_{g,j}\rangle}{M_j}$ are molar concentrations, and $M_j$ the molar mass, of gaseous species $j$.
We follow the same approach and use the above variables ($ \textbf{s}_L$) for model reduction.
Note that our choice of lifted variables is specific to the coupled porous--fluid system and differs from prior work (e.g., Swischuk et al.~\cite{swischuk2020learning}) due to the presence of heterogeneous reaction terms and interfacial heat transfer, which introduce additional coupling between gas and solid variables.
We consider the reactive source terms to be represented by a cubic operator, though this is an approximation.

\subsubsection{Low-dimensional state representation}

A projection-based ROM approximates the full-order system (Eq.~\ref{eq:states}) in a subspace with reduced
dimensionality $r << dn_{xy}$, spanned by a collection of $r$ basis vectors $\textbf{V} = [\textbf{v}_1 \;\; \textbf{v}_2 \dots \; \textbf{v}_r ] \in \mathbb{R}^{dn_{xy} \times r}$. One widely used approach to construct the basis $\textbf{V}$ is proper orthogonal decomposition (POD) \cite{guo2022bayesian,swischuk2020learning,peherstorferhar2016data}.

For a collection of known state snapshots $\textbf{q}_j:= \textbf{q}(t_j), \;\; j= 1, 2, \dots, k$, at $k$ time instances $t_1 < t_2 < \dots < t_k$, organized as $\textbf{Q}=[\textbf{q}_1 \;\; \textbf{q}_2 \; \dots \; \textbf{q}_k] \in \mathbb{R}^{dn_{xy} \times k}$, POD computes the singular value decomposition $\textbf{Q} = \bm{\Phi} \bm{\Sigma} \textbf{W}^T$, and uses the first $r$ columns of $\bm{\Phi}$ as the $r$-dimensional reduced basis $\textbf{V}$.

In classical projection-based model reduction, the reduced operators are obtained by projecting the full-order system
onto the space spanned by the POD basis vectors via, e.g., Galerkin projection \cite{peherstorferhar2016data}, with $\textbf{q} \approx \textbf{V} \hat{\textbf{q}}$:
\begin{equation}
    \diff{\hat{\textbf{q}}}{t} = \hat{\textbf{A}}(\bm{\eta})\hat{\textbf{q}} + \hat{\textbf{B}}(\bm{\eta})(\hat{\textbf{q}} \otimes \hat{\textbf{q}}) + \hat{\textbf{C}}(\bm{\eta})(\hat{\textbf{q}} \otimes \hat{\textbf{q}} \otimes \hat{\textbf{q}}) + \hat{\textbf{c}}(\bm{\eta}) + \dots\;,
    \label{eq:polyr}
\end{equation}
where $\hat{\textbf{A}}(\bm{\eta}) = \textbf{V}^T\textbf{A} \textbf{V} \in \mathbb{R}^{r \times r}$, 
$\hat{\textbf{B}}(\bm{\eta}) = \textbf{V}^T\textbf{B} (\textbf{V} \otimes \textbf{V}) \in \mathbb{R}^{r \times r^2}$, 
$\hat{\textbf{C}}(\bm{\eta}) = \textbf{V}^T\textbf{C} (\textbf{V} \otimes \textbf{V} \otimes \textbf{V}) \in \mathbb{R}^{r \times r^3}$, and 
$\hat{\textbf{c}}(\bm{\eta}) = \textbf{V}^T\textbf{c}  \in \mathbb{R}^{r}$ are ROM operators.
The key feature of this technique is that the ROM preserves the polynomial structure of the full-order model (Eq.~\ref{eq:poly}).

\subsubsection{Learning the operators by solving a least-squares problem}

The data-driven operator inference method attempts to learn the reduced-order operators ($\hat{\textbf{A}}$, $\hat{\textbf{B}}$, $\hat{\textbf{C}}$, and $\hat{\textbf{c}}$) utilizing a regression problem to find the operators that best match the snapshot data in a minimum-residual sense.
Such a non-intrusive approach eliminates the need for modifying the original governing equations or prior knowledge of the full-order model operators, which could be unfeasible for large systems of equations.

In this method the state snapshot data are first projected onto the POD subspace:
\begin{equation}
    \hat{\textbf{Q}}=\textbf{V}^T \textbf{Q} = \left[ \hat{\textbf{q}}_1(\bm{\eta}) \;\; \hat{\textbf{q}}_2(\bm{\eta}) \; \dots \; \hat{\textbf{q}}_k(\bm{\eta}) \right] \in \mathbb{R}^{r \times k}\;.
\end{equation}
The approximate time derivatives $ \dot{\hat{\textbf{Q}}}$ are computed from $\hat{\textbf{Q}}$ using a five-point finite-difference stencil:
\begin{equation}
    \dot{\hat{\textbf{Q}}} = \left[ \dot{\hat{\textbf{q}}}_1(\bm{\eta}) \;\; \dot{\hat{\textbf{q}}}_2(\bm{\eta}) \; \dots \; \dot{\hat{\textbf{q}}}_k(\bm{\eta}) \right] \in \mathbb{R}^{r \times k},
\end{equation}
where
\begin{equation}
\dot{\hat{\textbf{q}}}_j = \left(-\hat{\textbf{q}}_{j+2}+8\hat{\textbf{q}}_{j+1}-8\hat{\textbf{q}}_{j-1}+\hat{\textbf{q}}_{j-2} \right)/ \left(12\Delta t \right) \;,
\end{equation}
$j$ indicates the temporal index, and $\Delta t$ is the time-step size.
Then, operator inference solves the least squares problem:
\begin{multline}
    \min_{\hat{\textbf{A}}, \hat{\textbf{B}}, \hat{\textbf{C}}, \hat{\textbf{c}}} \Biggl\{ \sum_{j=1}^{k} \bigg\| \hat{\textbf{A}}(\bm{\eta})\hat{\textbf{q}}_j + \hat{\textbf{B}}(\bm{\eta})(\hat{\textbf{q}}_j \otimes \hat{\textbf{q}}_j) + \hat{\textbf{C}}(\bm{\eta})(\hat{\textbf{q}}_j \otimes \hat{\textbf{q}}_j \otimes \hat{\textbf{q}}_j) + \hat{\textbf{c}}(\bm{\eta}) - \dot{\hat{\textbf{q}}}_j \bigg\|^2_2 + \\  
    \bigg\|\bm{\Lambda}[\hat{\textbf{A}} \;\; \hat{\textbf{B}} \;\; \hat{\textbf{C}}  \;\; \hat{\textbf{c}}] \bigg\|^2_2 \Biggr\} \;,
%
\end{multline}
where $\boldsymbol{\Lambda} = \mathrm{diag}\left( 
\lambda_A \mathbf{I}_{p_A},\; \lambda_B \mathbf{I}_{p_B},\; 
\lambda_C \mathbf{I}_{p_C},\; \lambda_c \mathbf{I}_{p_c} \right)$; 
$\lambda_A$, $\lambda_B$, $\lambda_C$, and $\lambda_c$ are the regularization coefficients for the linear, quadratic, cubic, and constant operators, respectively; 
and $p_A=r$, $p_B=r\times(r+1)/2$, 
$p_C=r\times(r+1)\times(r+2)/6$, $p_c=1$ are the 
corresponding numbers of columns in each operator block.
We use Tikhonov regularization that allows different regularization strengths to be applied to each operator independently.

\subsubsection{Affine parametric model}

For parametric studies with a distribution of uncertain parameters $\bm{\eta}$, the parametric dependence of the operators should also be specified.
We assume that the operators exhibit the following affine decomposition with respect to  $\bm{\eta}$:
\begin{align}
    \textbf{c}(\bm{\eta}) &= \sum_{p=1}^{n_c} \theta_c^{(p)}(\bm{\eta}) \textbf{c}^{(p)} \;, \\
    \textbf{A}(\bm{\eta}) &= \sum_{p=1}^{n_A} \theta_A^{(p)}(\bm{\eta}) \textbf{A}^{(p)} \;, \\
    \textbf{B}(\bm{\eta}) &= \sum_{p=1}^{n_B} \theta_B^{(p)}(\bm{\eta}) \textbf{B}^{(p)} \;, \\
    \textbf{C}(\bm{\eta}) &= \sum_{p=1}^{n_C} \theta_C^{(p)}(\bm{\eta}) \textbf{C}^{(p)} \;,
\end{align}
where $\theta_c^{(p)}, \; \theta_A^{(p)}, \; \theta_B^{(p)}, \; \theta_C^{(p)}$ are known scalar-valued coefficient functions;
$n_c$, $n_A$, $n_B$, and $n_C$ are the respective number of terms.
The operators $\textbf{c}^{(p)}, \; \textbf{A}^{(p)}, \; \textbf{B}^{(p)}, \; \textbf{C}^{(p)}$ are independent of $\bm{\eta}$. Such a structure may occur naturally in the governing equations.
For example, the term $h_{fs} A_{fs} \left(\langle T \rangle - \langle T^s \rangle\right)$ in Eq.~\eqref{eq:4} exhibits this structure: the parameter $h_{fs}$ enters only as a multiplicative
factor, so with $\boldsymbol{\eta} = \left(h_{fs}\right)$ every operator contribution arising from
this term carries the same coefficient function, the identity map
$\theta^{(p)}(\boldsymbol{\eta}) = h_{fs}$, acting on parameter-independent operators
$\mathbf{A}^{(p)}$, $\mathbf{B}^{(p)}$, etc.

A key advantage of embedding the affine parametric structure directly into the operator inference regression problem is that the minimum number of parameter samples required for well-posedness is determined by the number of terms in the affine decomposition of the operators in the above equations, rather than by the dimensionality of the parameter space itself~\cite{mcquarrie2023nonintrusive}.
This means the framework can, in principle, handle multi-dimensional parameter spaces without an exponential growth in the number of required full-order model evaluations, provided the affine parametric structure of the governing equations is preserved.

To learn the above operators, the procedure in the previous sections is modified to include the affine parametric model.
Supposing we can sample the solution of the full-order model at $M$ parameter values $\{\bm{\eta}_i\}_{i=1}^M \subset \mathcal{P}$, the least squares problem then becomes:
\begin{multline}
    \min_{\hat{\textbf{A}}, \hat{\textbf{B}}, \hat{\textbf{C}}, \hat{\textbf{c}}} \Biggl\{ \sum_{i=1}^{M}\sum_{j=1}^{k} \bigg\| \hat{\textbf{A}}(\bm{\eta}_i)\hat{\textbf{q}}_j + \hat{\textbf{B}}(\bm{\eta}_i)(\hat{\textbf{q}}_j \otimes \hat{\textbf{q}}_j) + \hat{\textbf{C}}(\bm{\eta}_i)(\hat{\textbf{q}}_j \otimes \hat{\textbf{q}}_j \otimes \hat{\textbf{q}}_j) + \hat{\textbf{c}}(\bm{\eta}_i) - \dot{\hat{\textbf{q}}}_j(\bm{\eta}_i) \bigg\|^2_2  \\ 
    + \bigg\| \bm{\Lambda}[\hat{\textbf{A}} \;\; \hat{\textbf{B}} \;\; \hat{\textbf{C}}  \;\; \hat{\textbf{c}}] \bigg\|^2_2 \Biggl\}
\end{multline}
where $\dot{\hat{\mathbf{q}}}_j(\boldsymbol{\eta}_i)$ and the POD basis are derived from the updated
state snapshot data matrix
\begin{equation*}
\left[ \mathbf{Q}(\boldsymbol{\eta}_1) \; \mathbf{Q}(\boldsymbol{\eta}_2) \; \dots \;
\mathbf{Q}(\boldsymbol{\eta}_M) \right] \in \mathbb{R}^{dn_{xy} \times Mk} \;.
\end{equation*}
In practice the least squares problem above is solved in linear form for the unknown operators,
with data matrix $\mathbf{D} \in \mathbb{R}^{Mk \times (n_A r + p_B + p_C + 1)}$ whose rows are
\begin{equation*}
\left[ \theta_A^{(1)}(\boldsymbol{\eta}_i)\hat{\mathbf{q}}_j^T \;\; \dots \;\;
\theta_A^{(n_A)}(\boldsymbol{\eta}_i)\hat{\mathbf{q}}_j^T \;\;
(\hat{\mathbf{q}}_j \otimes \hat{\mathbf{q}}_j)^T \;\;
(\hat{\mathbf{q}}_j \otimes \hat{\mathbf{q}}_j \otimes \hat{\mathbf{q}}_j)^T \;\; 1 \right] \;,
\end{equation*}
for $i = 1, \dots, M$ and $j = 1, \dots, k$; the conditioning of $\mathbf{D}$ governs the
well-posedness of the regression.

\section{Results and discussion}
This section presents the results of reduced-order modeling and uncertainty quantification for two applications: emission of buoyant reacting plumes from the surface of a heated solid and an ablation experiment under atmospheric entry and high-enthalpy flow conditions.
For the latter problem, we first describe the results of the full-order model applied to the ablation experiment, followed by the reduced-order modeling outcomes.

\begin{figure}[htbp]
\centering
\includegraphics[width=.5\textwidth]{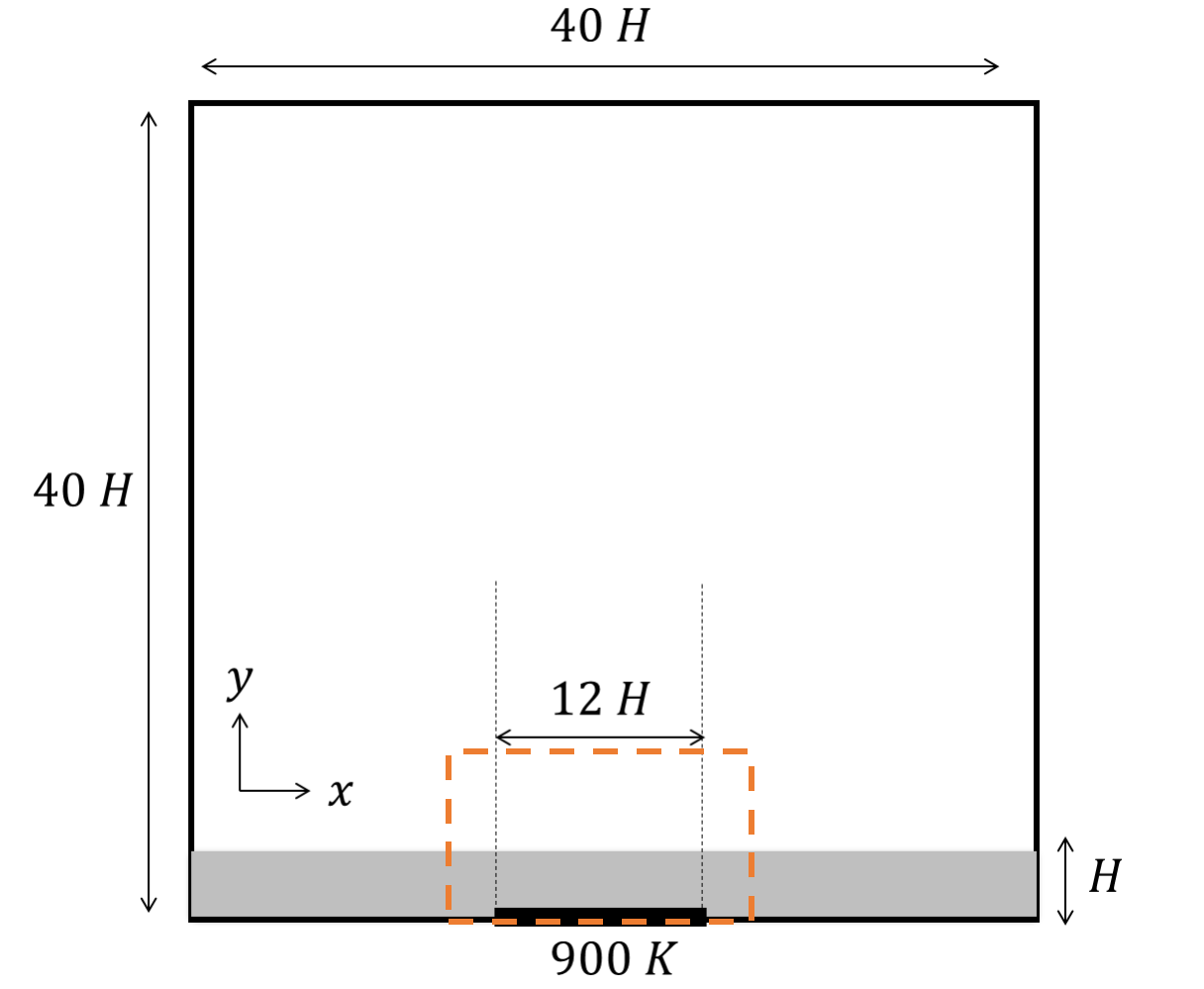}
\caption{Computational domain for the buoyant reacting plumes simulation. The portion of the domain used for reduced-order modeling is indicated by the orange dashed line and has dimensions of $14.4H$ by $10H$.}
\label{fig:plume_domain}
\end{figure}

\subsection{Solid-fuel combustion in quiescent air}
The formation of buoyant reacting plumes is a situation that can occur in the burning of solid fuels or natural fires when a porous layer is heated from below in the absence of a forced flow. Our previous work \cite{behnoudfar2025single} obtained two-dimensional simulations of this process using the full-order model described in Section~\ref{subsec:gequation}.
We utilize those data to learn a cubic ROM for the process using the following lifted state variables:
\begin{equation*}
    \left[ \langle u_x \rangle \;\;\langle u_y \rangle \;\; \langle T \rangle \;\;\langle T_s \rangle \;\; \frac{1}{\phi\langle \rho \rangle^f} \;\; \langle c_{CH_4} \rangle \;\;\langle c_{O_2}\rangle\;\;\langle c_{CO_2}\rangle\;\;\langle c_{CO}\rangle \;\;\langle c_{H_2O}\rangle \;\;\langle c_{N_2}\rangle\;\;\langle Y_{\text{wood}} \rangle \;\;\langle Y_{\text{char}}\rangle \right] \;.
\end{equation*}
The system involves three reactions: in the solid phase, wood decomposes into char and other gaseous species, and, in the gas phase, two oxidation reactions occur for \ce{CH4} and \ce{CO}.
Training data are generated by running full-order simulations over a time span of \SI{0.5}{\second}, with the state saved at 100 uniformly spaced instants ($\Delta t = \SI{5}{\milli\second}$).
The analysis is done using a portion of the simulated domain as shown in Figure~\ref{fig:plume_domain}, resulting in a grid size of $n_{xy}=14400$.
Due to the large differences in the variable scales, the data are pre-processed to scale each variable to the interval $[-1, 1]$.

\subsubsection{Non-parametric model performance \label{subsubsec:non-parametric}}

\begin{figure}[htbp]
\centering
\begin{subfigure}[t]{.48\textwidth}%
\centering\captionsetup{width=.8\linewidth}%
\includegraphics[width=\linewidth]{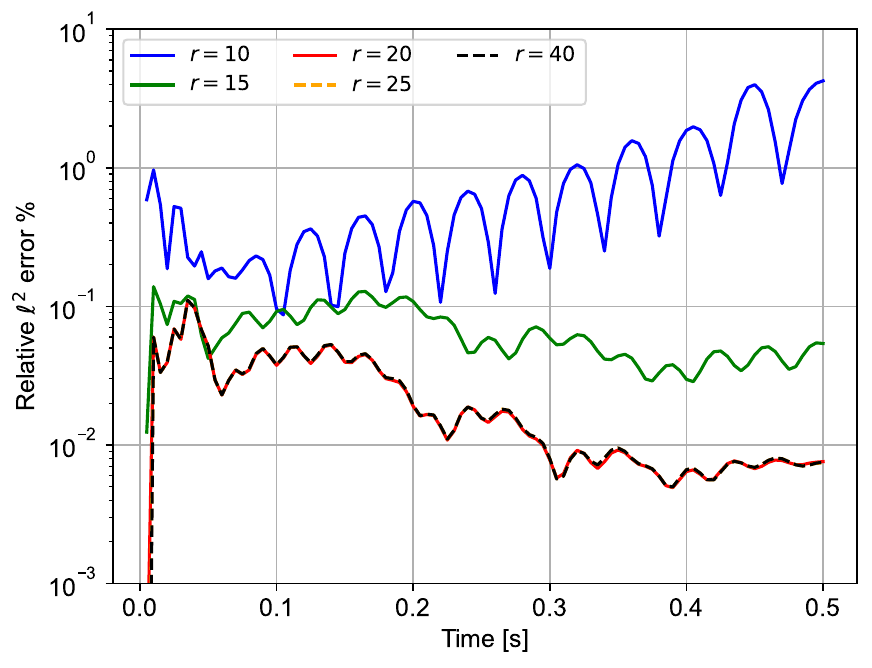}%
\caption{Relative $\ell^2$-norm error error versus time}
\label{fig:error}
\end{subfigure}
~
\begin{subfigure}[t]{.48\textwidth}%
    \centering\captionsetup{width=.8\linewidth}%
\includegraphics[width=\linewidth]{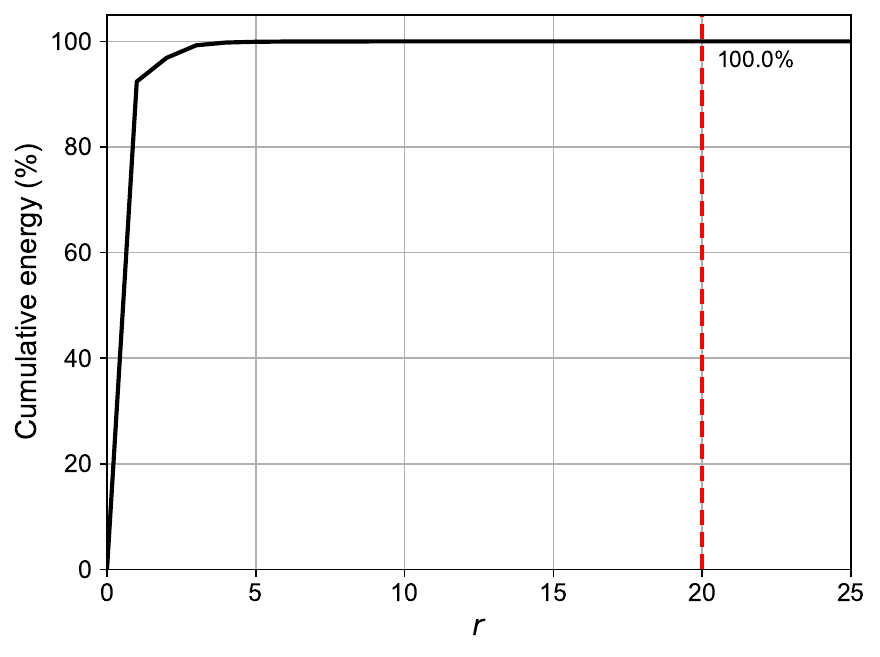}%
\caption{Cumulative energy versus basis size ($r$)}
\label{fig:cumulative_energy}
\end{subfigure}

\caption{Performance of non-parametric ROM for reacting buoyant plume, for permeability associated with the initial porosity of $\phi_0$ = 0.875.}
\label{fig:plume_error}
\end{figure}

Figure~\ref{fig:plume_error} shows the ROM performance over time for the permeability associated with the initial porosity of $\phi_0 = 0.875$. The performance is measured by relative $\ell^2$-norm error defined as
\begin{equation*}
    \mathcal{E}_{j,\bm{\eta}} = \frac{\big\|\textbf{V} \hat{\textbf{q}}(t_j; \bm{\eta}) - \textbf{q}(t_j; \bm{\eta}) \big\|_2}{\big\| \textbf{q}(t_j; \bm{\eta}) \big\|_2} \;.
\end{equation*}
The error decreases with increasing basis size up to $r=20$, beyond which further increases yield little improvement, indicating that the reduced data contain sufficient information to represent the system.
This corresponds to cumulative energy of essentially 100\%, as seen in Figure~\ref{fig:cumulative_energy}.
Cumulative energy measures the amount of singular value energy captured by the basis, $\sum_{i=1}^{r} \sigma_i^2/\sum_{i=1}^{k} \sigma_i^2$, where $\sigma_i$ are the diagonal elements of $\bm{\Sigma}$.
The rapid convergence of the cumulative energy to 100\% indicates that the snapshot data are well-represented by a low-dimensional subspace of this type, which is a favorable condition for reduced-order modeling.
The initial spike in ROM error has two main causes: the FOM fields show an abrupt rise right at the start that is hard for the ROM (trained on derivatives) to capture, and derivative data near time boundaries are less accurate due to one-sided differences.

\begin{figure}[htbp]
\centering
\includegraphics[width=.5\textwidth]{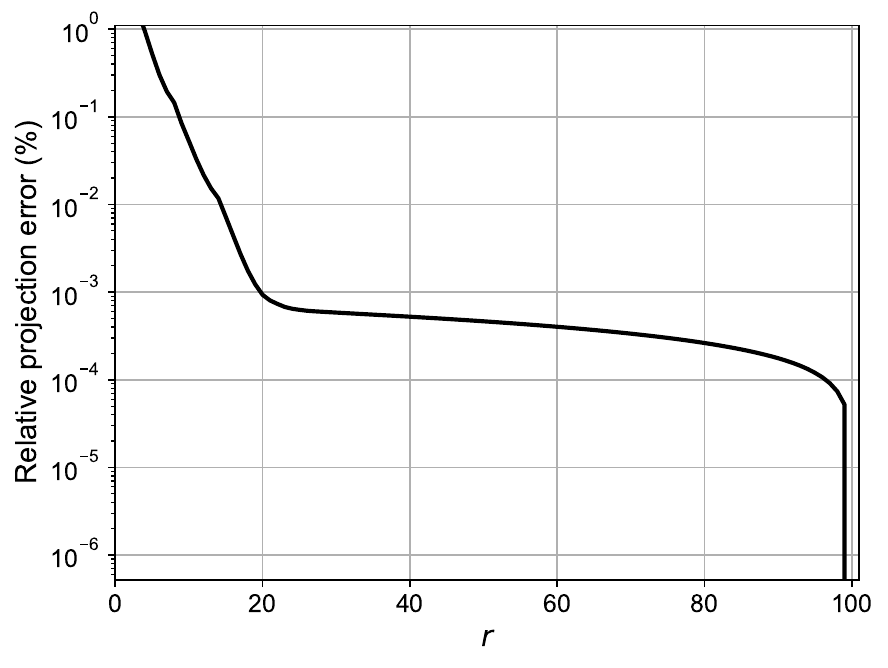}
\caption{Relative projection error versus basis size for the reacting buoyant plume case; the error is reported as a percentage.
The sharp drop as $r$ approaches 100 occurs because the basis then spans all $k=100$ snapshots, so the
projection becomes exact to within round-off.}
\label{fig:projection_error}
\end{figure}

To help further explain the observed changes in error with basis size,
Figure~\ref{fig:projection_error} shows the relative error in projection, represented by the difference between $\mathbf{V}\hat{\mathbf{q}}^{\text{proj}}$ and the snapshot data $\mathbf{q}$, 
where $\hat{\mathbf{q}}^{\text{proj}} = \mathbf{V}^T \mathbf{q}$.
We compute the projection error as
\begin{equation}
\frac{\big\|\textbf{V} \textbf{V}^T \textbf{Q}(\bm{\eta}) - \textbf{Q}( \bm{\eta}) \big\|_F}{\big\| \textbf{Q}(\bm{\eta}) \big\|_F} \;,
\end{equation}
where $\big\| \cdot \big\|_F$ is the Frobenius norm.
The systematic decrease in relative error as the basis size increases from $r = 10$ to $r = 20$ seen in Figure~\ref{fig:error} reflects the projection, or truncation, error, while the plateau observed beyond $r = 20$ indicates that the remaining error is dominated by regression error from the learned operators rather than by an insufficient reduced subspace.


\begin{figure}[htbp]
\centering
\begin{subfigure}[t]{.46\textwidth}%
\centering\captionsetup{width=\linewidth}%
\includegraphics[width=\linewidth]{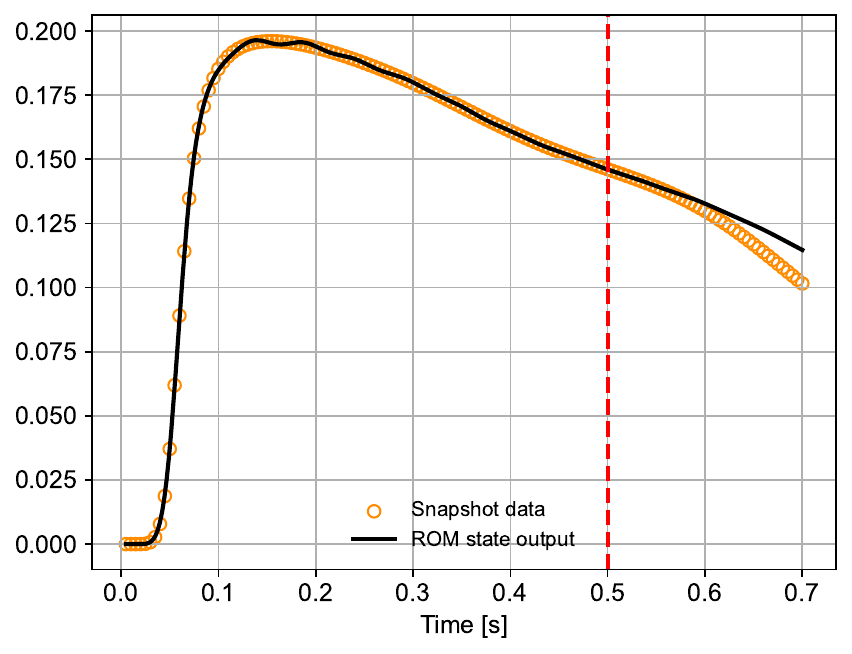}%
\caption{$r=20$}
\label{}
\end{subfigure}
~
\begin{subfigure}[t]{.48\textwidth}%
\centering\captionsetup{width=\linewidth}%
\includegraphics[width=\linewidth]{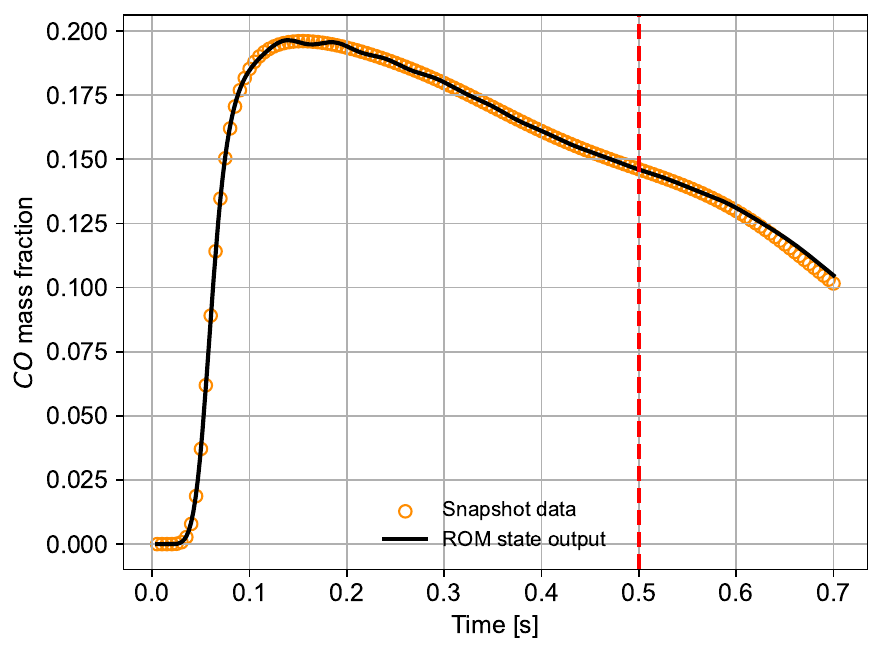}%
\caption{$r=40$}
\label{}
\end{subfigure}

\caption{Effect of basis size on ROM extrapolation accuracy in CO mass fraction for the reacting buoyant plume case.}
\label{fig:basis_size_effect}
\end{figure}

However, while using a basis size beyond $r=20$ provides minimal benefit inside the training window, a larger basis size can improve extrapolated predictions past this point.
Figure~\ref{fig:basis_size_effect} compares predictions of \ce{CO} mass fraction using $r=20$ and $r=40$, showing that doubling the basis size visibly improves the predictions past $t = \SI{0.5}{\second}$.
For this reason, we chose $r=40$ for the remaining analysis.

\begin{figure}[htbp]
\centering
\begin{subfigure}[t]{.4\textwidth}%
\centering\captionsetup{width=.8\linewidth}%
\includegraphics[width=\linewidth]{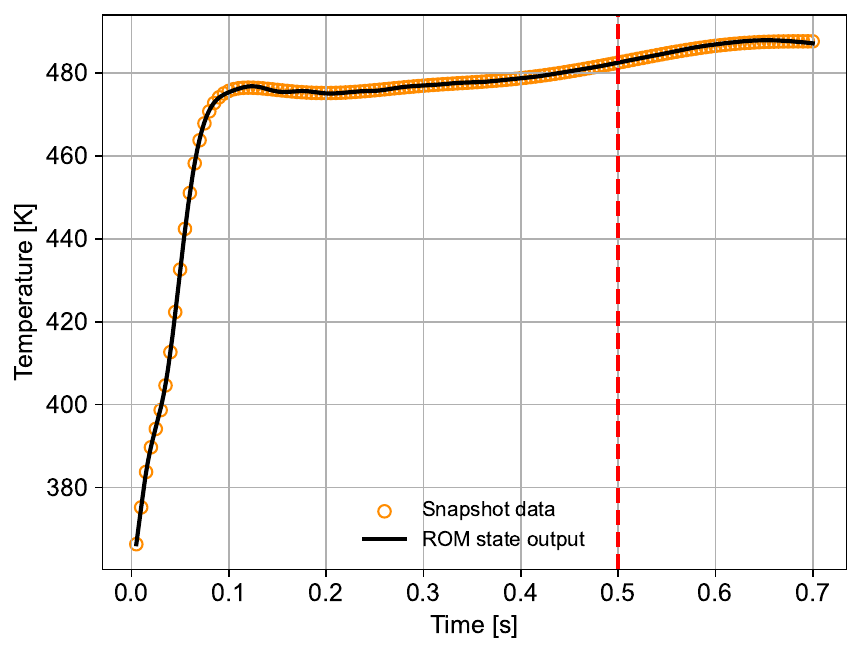}%
\caption{Gas temperature}
\label{fig:/T_plume}
\end{subfigure}
~
\begin{subfigure}[t]{.4\textwidth}%
    \centering\captionsetup{width=.8\linewidth}%
\includegraphics[width=\linewidth]{figs/co_plume.pdf}%
\caption{$\ce{CO}$ mass fraction}
\label{fig:co_plume}
\end{subfigure}

\begin{subfigure}[t]{.4\textwidth}%
    \centering\captionsetup{width=.8\linewidth}%
\includegraphics[width=\linewidth]{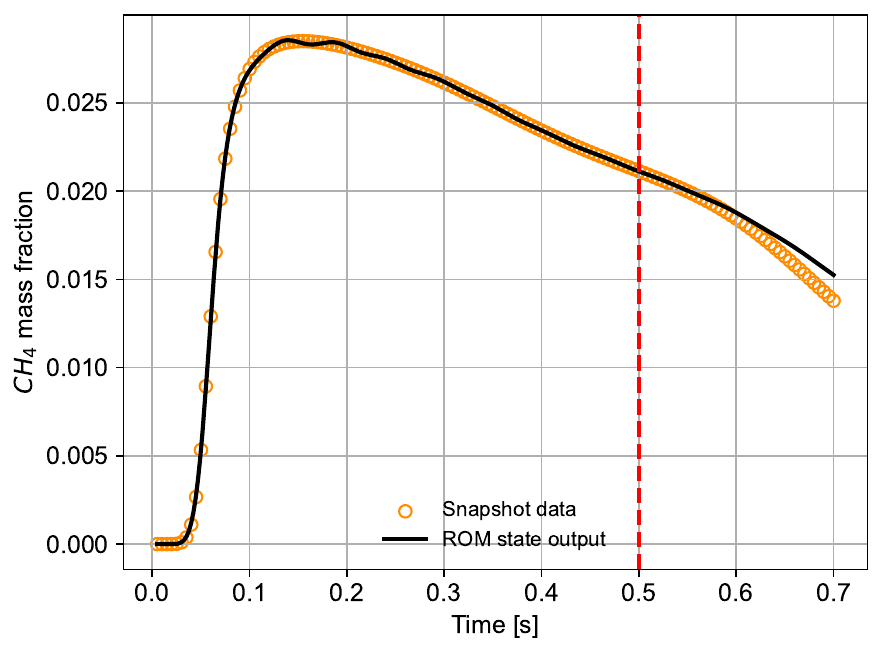}%
\caption{$\ce{CH4}$ mass fraction}
\label{fig:ch4_plume}
\end{subfigure}
~
\begin{subfigure}[t]{.4\textwidth}%
    \centering\captionsetup{width=.8\linewidth}%
\includegraphics[width=\linewidth]{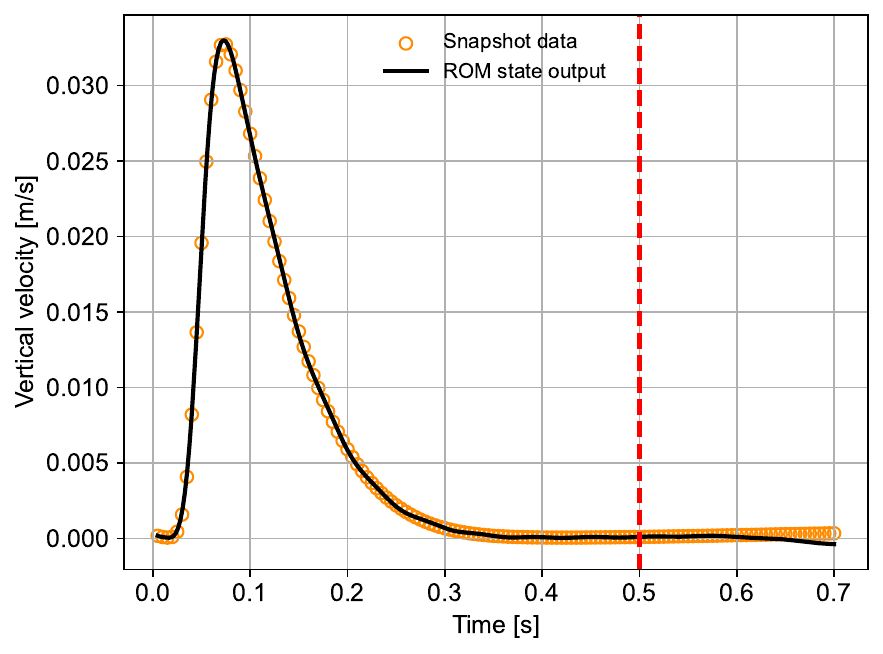}%
\caption{Gas vertical velocity}
\label{fig:vy_plume}
\end{subfigure}

\caption{Evolution of quantities of interest at the center point on the interface in the reacting buoyant plume, with $r=40$ and $\boldsymbol{\Lambda} = \lambda \mathbf{I}$ with $\lambda = \num{e4}$. The dashed line indicates the end of the training data.}
\label{fig:plume_QoI}
\end{figure}

\begin{figure}[htbp]
\centering
\begin{subfigure}[t]{.4\textwidth}%
\centering\captionsetup{width=.8\linewidth}%
\includegraphics[width=\linewidth]{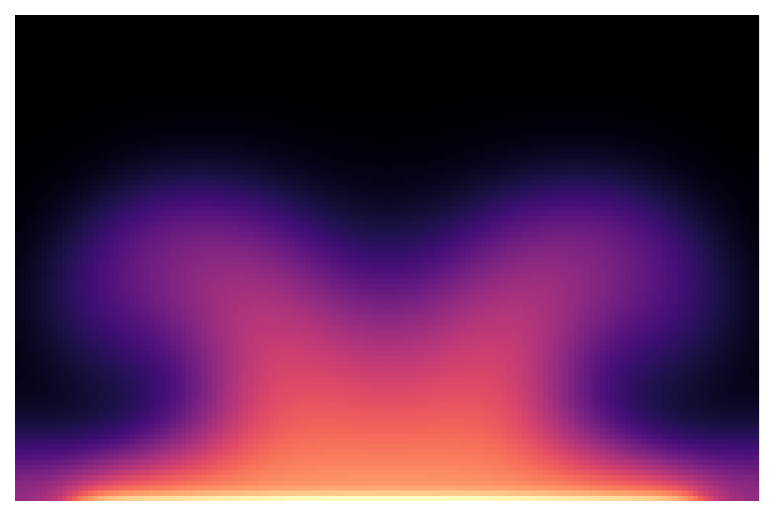}%
\caption{True gas temperature}
\label{fig:T_plume_true}
\end{subfigure}
~
\begin{subfigure}[t]{.475\textwidth}%
    \captionsetup{width=.8\linewidth}%
\includegraphics[width=\linewidth]{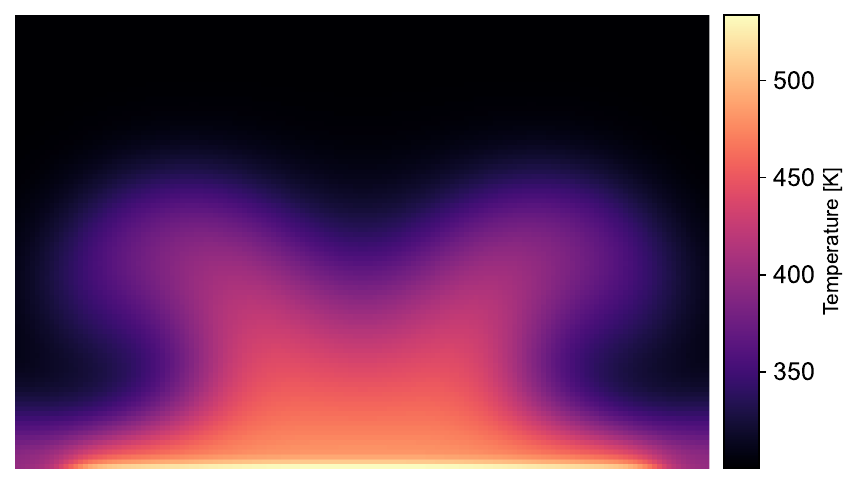}%
\caption{ROM-predicted gas temperature}
\label{fig:T_plume_rom}
\end{subfigure}

\begin{subfigure}[t]{.4\textwidth}%
    \centering\captionsetup{width=.8\linewidth}%
\includegraphics[width=\linewidth]{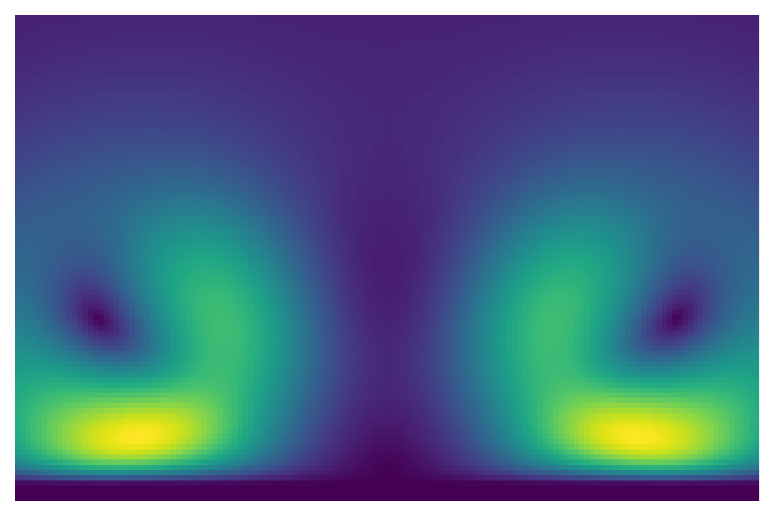}%
\caption{True velocity magnitude}
\label{fig:velocity_plume_true}
\end{subfigure}
~
\begin{subfigure}[t]{.475\textwidth}%
    \captionsetup{width=.8\linewidth}%
\includegraphics[width=\linewidth]{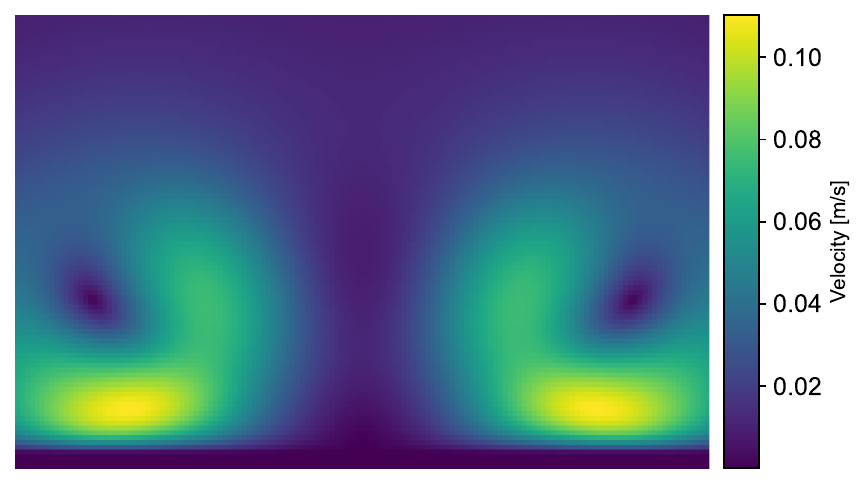}%
\caption{ROM-predicted velocity magnitude}
\label{fig:velocity_plume_rom}
\end{subfigure}

\caption{Non-parametric model predictions at the last time step of training data for the reacting buoyant plume.}
\label{fig:plume_2d}
\end{figure}

Figure~\ref{fig:plume_QoI} compares the quantities of interest, including the gas temperature, mass fractions of \ce{CH4} and \ce{CO}, and vertical velocity, predicted by the ROM and full-order models at a monitor point at the center of the interface (specified in Figure~\ref{fig:plume_domain}).
The ROM predictions agree well with the training data; the results also demonstrate the ability of the model to extrapolate beyond the training data.  Figure~\ref{fig:plume_2d} shows the non-parametric model predictions at the last time step of training data, demonstrating that the model successfully captures the spatial variations.
The full-order model requires an average wall-clock time of $\sim$284 seconds, running in parallel across four CPU cores with 16 GB of memory on a 20-core Intel Xeon E5 system with 128 GB of RAM.
The ROM at $r=40$ completes the same integration in $\sim$3.0 seconds, with a computational speedup of approximately $95\times$.

\begin{figure}[htbp]
\centering
\begin{subfigure}[t]{.8\textwidth}%
\captionsetup{width=1\linewidth}%
\includegraphics[width=\linewidth]{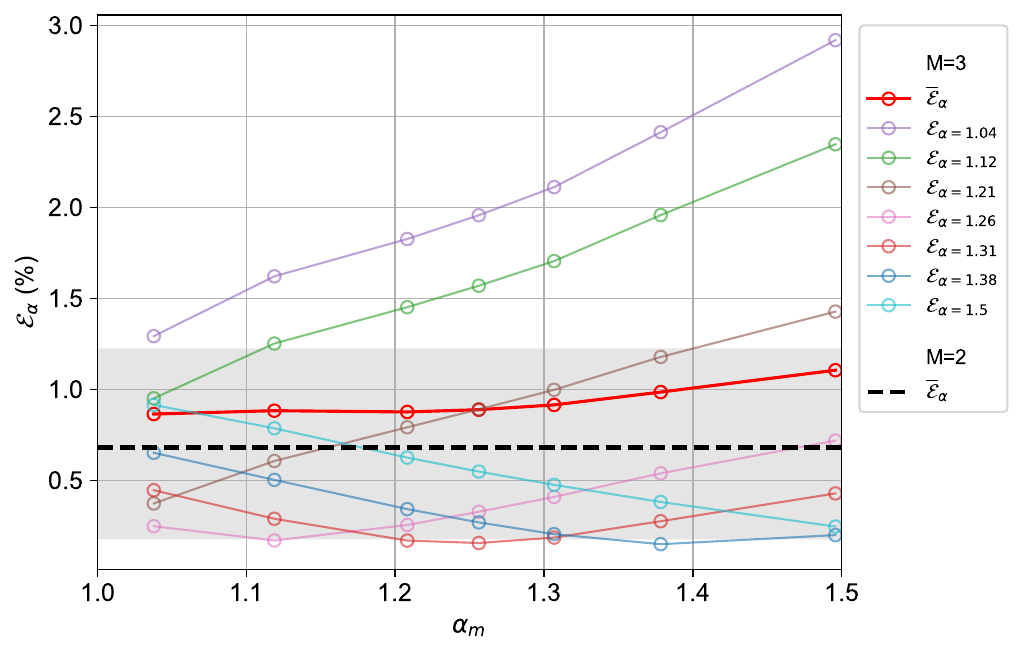}%
\caption{Relative error ($\mathcal{E}_{\alpha}$) with varying middle point parameter ($\alpha_m$) location evaluated at different $\alpha$ values using $k=50$ time snapshots (the error is averaged over time). The shaded area represents the range of variations in the error of the two-parameter model.}
\label{fig:errorvsmiddlepoint}
\end{subfigure}

\begin{subfigure}[t]{.8\textwidth}%
    \captionsetup{width=1\linewidth}%
\includegraphics[width=\linewidth]{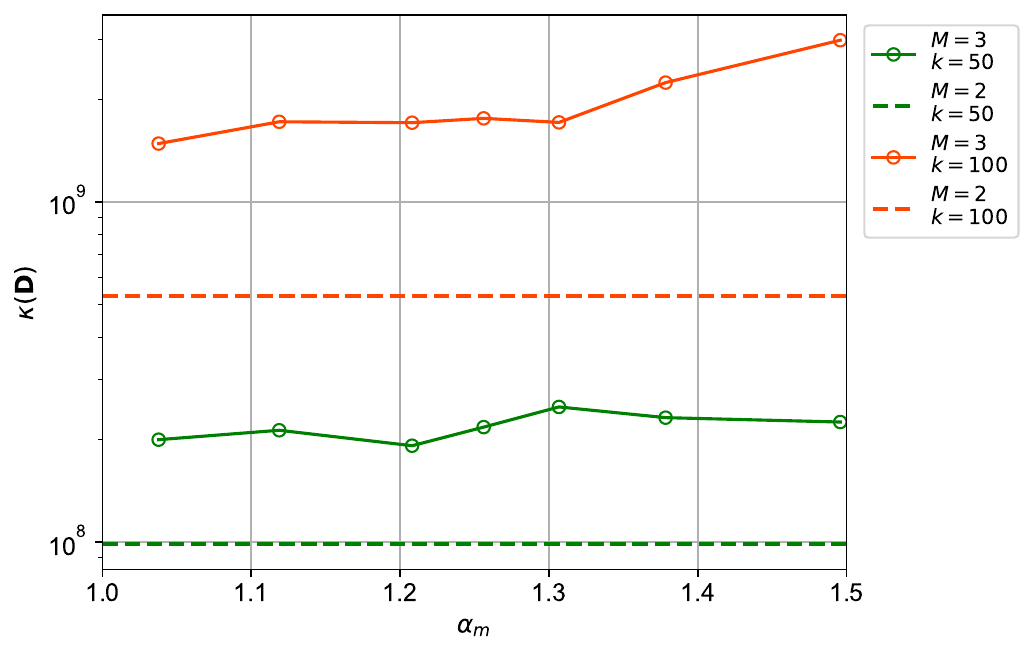}%
\caption{Condition number with varying middle point location for different parameter set sizes and number of time snapshots. Note the semilogarithmic scale on the vertical axis.}
\label{fig:cond}
\end{subfigure}

\caption{Sensitivity of the ROM to the number and location of parameter samples.}
\label{fig:uq_error_cond}
\end{figure}

\subsubsection{Parametric model performance}
\label{subsec:parametric_model}

Returning to the primary objective of this study, we construct a parametric model with an affine linear component to quantify the uncertainty in permeability ($\gamma$), where $\textbf{K} = \gamma(\phi_0) \textbf{I}$ and $\phi_0$ is initial porosity.
The ROM has the same cubic form as in Eq.~\eqref{eq:polyr}, with the difference that the linear
operator admits a two-term affine decomposition, $\mathbf{A}(\boldsymbol{\eta}) =
\gamma^{-1}\mathbf{A}^{(1)} + \mathbf{A}^{(2)}$, that is, $n_A = 2$ with
$\theta_A^{(1)} = \gamma^{-1}$ and $\theta_A^{(2)} = 1$, while $n_c = n_B = n_C = 1$ with
$\theta_c^{(1)} = \theta_B^{(1)} = \theta_C^{(1)} = 1$; the inverse dependence on $\gamma$ follows
from the drag force term $\mathbf{f} = - \mu \mathbf{K}^{-1}[\mathbf{I}+\text{Re}_p \mathbf{F}_0]
\cdot \langle\mathbf{u}\rangle$ in Eq.~\eqref{eq:3} with $\mathbf{K} = \gamma \mathbf{I}$.
We generate data at each of the parameter realizations $\bm{\eta} = \left(\gamma_{\phi_0}\right)$ for 
$\gamma \in 10^{-9} \times \{ $\numlist[list-final-separator={, }]{1.045; 1.084; 1.126; 1.169; 1.215; 1.262; 1.313; 1.366; 1.440; 1.563; 1.700; 1.853; 2.056}$\}$,
with the range chosen based on previous observations~\cite{behnoudfar2025single}.
This range corresponds to 13 values of the initial porosity, $\phi_0 \in \{$\numlist[list-final-separator={, }]{0.825;0.828;0.831;
0.834;0.837;0.840;0.843;0.846;0.850;0.856;0.862;0.868;0.875}$\}$.

Notably, a finer parameter grid does not necessarily yield a better learned ROM and may even degrade the model quality. Figure~\ref{fig:uq_error_cond} shows the result of an analysis on the sensitivity of the ROM to the number and location of parameter samples.
We compare the performance of the built models using the parameter set sizes of $M=2$ and 3.
Figure~\ref{fig:errorvsmiddlepoint} shows that for $M=3$ the mean model error ($\overline{\mathcal{E}}_{\alpha}$), averaged over the evaluation points $\alpha = \frac{\gamma}{\gamma_1} \in [1,\;\frac{\gamma_3}{\gamma_1}]$, increases with the distance between the middle and the first parameter.
The middle parameter location is defined as $\alpha_m=\frac{\gamma_2}{\gamma_1}$, where $\mathcal{P} = \{\gamma_1,\; \gamma_2,\; \gamma_3\}$.
Here, $\gamma_1$ corresponds to $\phi_0 = 0.825$.
In a local sense, the model performs better as $\alpha_m$ becomes closer to the evaluation point.
Thus, for $\alpha > 1.31$, $\mathcal{E}_{\alpha}$ decreases with $\alpha_m$ while for $\alpha < 1.26$, it increases with $\alpha_m$.
Interestingly, the model built using two parameter samples ($M=2$) outperforms the models constructed with three parameter samples ($M=3$).
If the middle data point is not informative---that is, if the dynamics are not sufficiently diverse---it can lead to rank deficiencies in the data matrix $\textbf{D}$.
This is consistent with the larger condition numbers $\kappa(\textbf{D})$ for $M=3$ cases as shown in Figure~\ref{fig:cond}.
Doubling the number of time snapshots to $k=100$ increases the condition number by approximately an order of magnitude.

\begin{figure}[htbp]
\centering
\includegraphics[width=.6\textwidth]{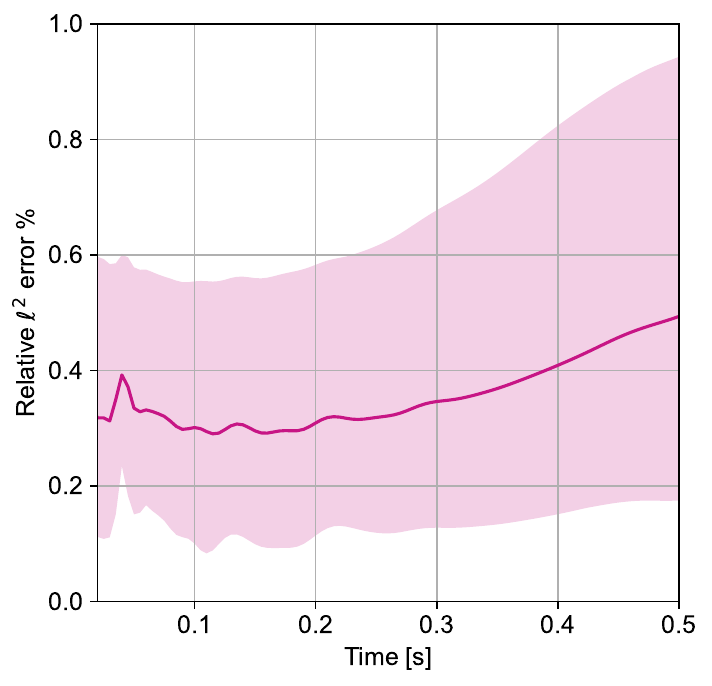}
\caption{Relative error of parametric ROM with time (representing the norm of error contribution from all variables), averaged over the test data. The solid line denotes the mean error and the shaded area the standard deviation.}
\label{fig:test_error_uq}
\end{figure}

\begin{figure}[htbp]
\centering
\begin{subfigure}[t]{.4\textwidth}%
\centering\captionsetup{width=.8\linewidth}%
\includegraphics[width=\linewidth]{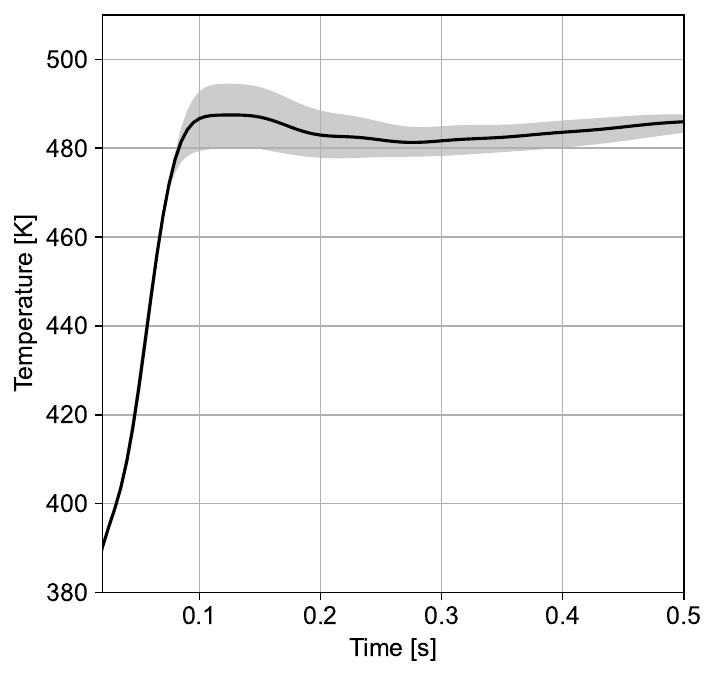}%
\caption{Gas temperature ($T$)}
\label{fig:Tuq_plume}
\end{subfigure}
~
\begin{subfigure}[t]{.4\textwidth}%
    \centering\captionsetup{width=.8\linewidth}%
\includegraphics[width=\linewidth]{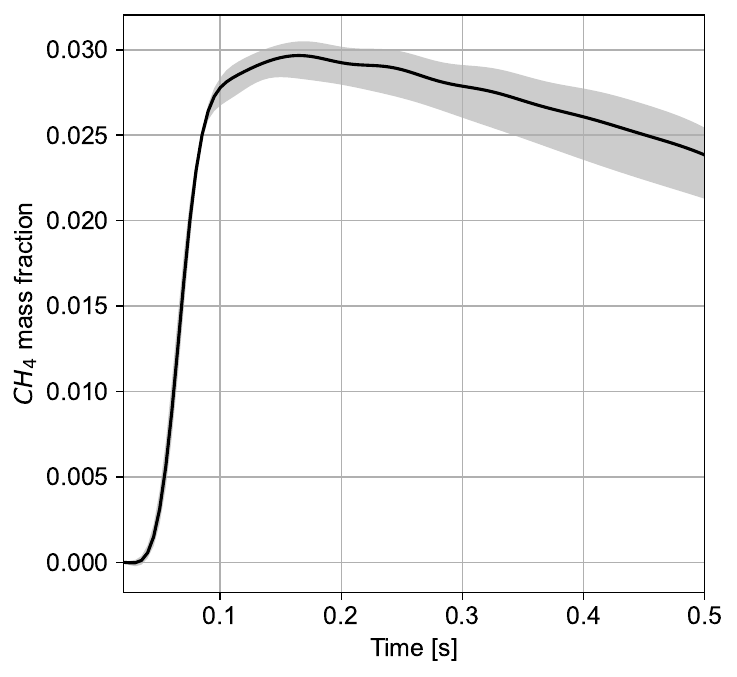}%
\caption{$\ce{CH4}$ mass fraction ($Y_{\ce{CH4}}$)}
\label{fig:ch4uq_plume}
\end{subfigure}

\begin{subfigure}[t]{.4\textwidth}%
    \centering\captionsetup{width=.8\linewidth}%
\includegraphics[width=\linewidth]{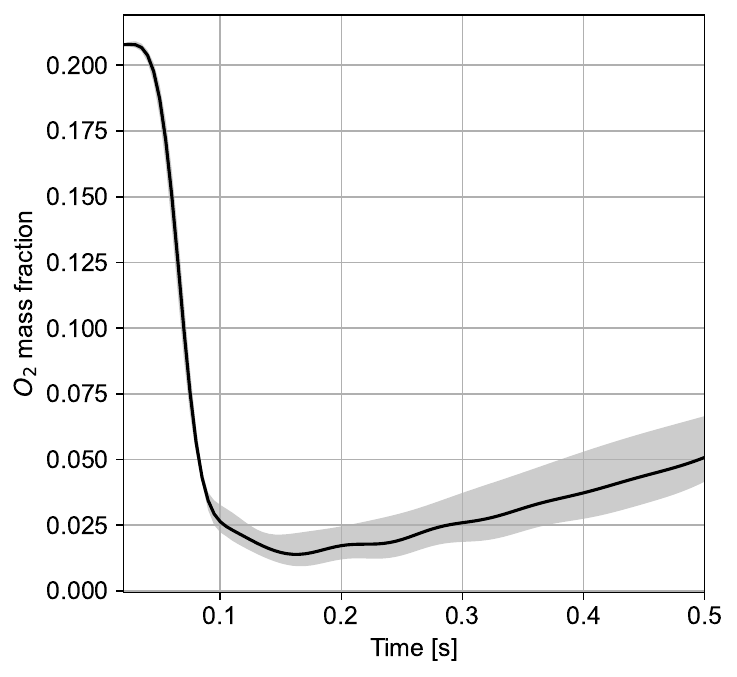}%
\caption{$\ce{O2}$ mass fraction ($Y_{\ce{O2}}$)}
\label{fig:o2uq_plume}
\end{subfigure}
~
\begin{subfigure}[t]{.4\textwidth}%
    \centering\captionsetup{width=.8\linewidth}%
\includegraphics[width=\linewidth]{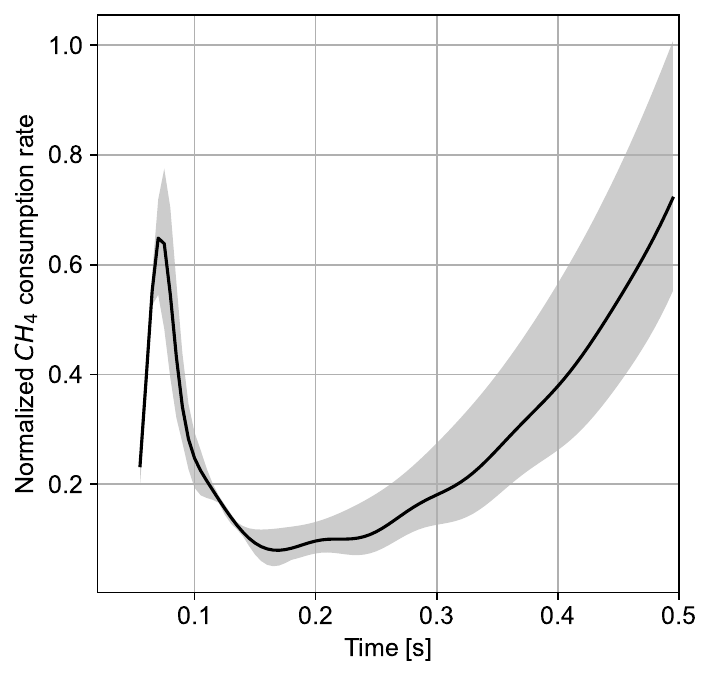}%
\caption{Rate of \ce{CH4} oxidation ($\dot{\omega}_{\ce{CH4}}$)}
\label{fig:rate}
\end{subfigure}

\caption{Evolution of uncertain quantities of interest at the center point on the interface with $r=40$, $\boldsymbol{\Lambda} = 1.2\times 10^{4}\,\mathbf{I}$, $k=50$ and $N=150$. The shaded area indicates the standard deviation.}
\label{fig:plumeUQ_QoI}
\end{figure}

Following the above observations, we construct a parametric ROM by combining two sub-models:
\begin{align*}
    f_{\mathcal{P}_1}(\textbf{q};\gamma),\; \mathcal{P}_1 &= \{\gamma_{\phi_0=0.825},\; \gamma_{\phi_0=0.85}\}: \qquad \gamma \in [\gamma_{\phi_0=0.825},\; \gamma_{\phi_0=0.85}] \\
    f_{\mathcal{P}_2}(\textbf{q};\gamma),\; \mathcal{P}_2 &= \{\gamma_{\phi_0=0.85},\; \gamma_{\phi_0=0.875}\}: \qquad \gamma \in [\gamma_{\phi_0=0.85},\; \gamma_{\phi_0=0.875}] \;.
\end{align*}
For a Monte Carlo-like analysis, the model is sampled with a finite set of $N$ values of the affine
parameter $\gamma^{-1}$, drawn from a normal distribution centered on the midpoint of the parameter
range with a standard deviation of one sixth of that range, so that the range corresponds to
$\mu \pm 3\sigma$:
\begin{equation*}
\gamma^{-1} \sim \mathcal{N}\!\left(\frac{{\gamma}^{-1}_{\phi_0=0.875} + {\gamma}^{-1}_{\phi_0=0.825}}{2},\;
\left[\frac{{\gamma}^{-1}_{\phi_0=0.825} - {\gamma}^{-1}_{\phi_0=0.875}}{6}\right]^{2}\right) \;,
\end{equation*}
with samples falling outside $[{\gamma}^{-1}_{\phi_0=0.875},\; {\gamma}^{-1}_{\phi_0=0.825}]$
clipped to the nearest endpoint.
For each sub-model, the basis is computed locally from the snapshots spanning the parameter range associated with the sub-model.
Figure~\ref{fig:test_error_uq} presents the resulting model error evaluated on the test data, which comprises 10 of the aforementioned parameter realizations, excluding those in $\mathcal{P}_1$ and $\mathcal{P}_2$.
The error remains below 1\% throughout and gradually increases toward the end, mainly due to one-sided data near the time boundaries, which makes it harder to capture the derivatives accurately.
Figure~\ref{fig:plumeUQ_QoI} shows the forward propagation of uncertainty through the system.
Quantifying the uncertainty in the rate of combustion reactions could be an important outcome of such an analysis.
As seen in Figure~\ref{fig:rate}, the rate of \ce{CH4} oxidation can vary up to 100\% later in the simulation, which can have implications for ignition processes and fire propagation.
The broader uncertainty bounds observed at later times across multiple quantities of interest can be partly attributed to the stronger dynamics and higher reaction rates at these stages, which amplify the sensitivity of the coupled reacting flow--porous solid system to variations in the uncertain parameters.


\subsection{Ablation in a plasma wind tunnel}

Here, we first report the results of the numerical analysis of the ablation experiment as conducted inside a plasmatron facility \cite{helber2015microstructure}, using the physics-based model described in Section~\ref{subsec:gequation}.
The following subsection discusses reduced-order modeling and uncertainty quantification based on operator inference. 

\begin{figure}[htbp]
\centering
\includegraphics[width=.7\textwidth]{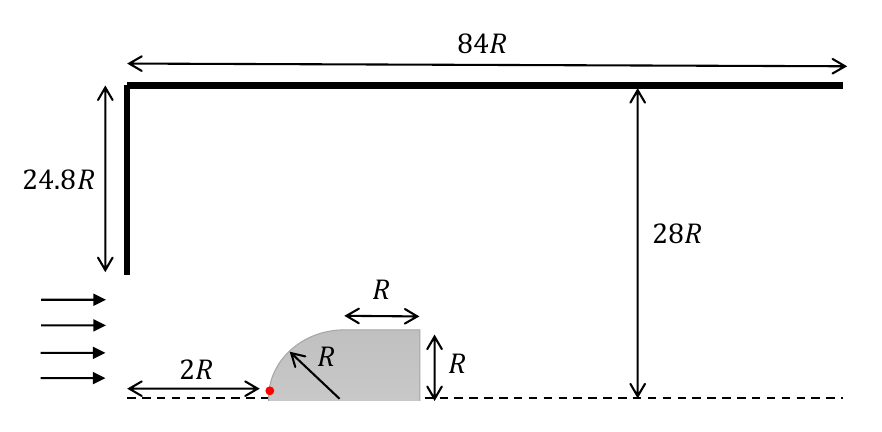}
\caption{Computational domain for the ablation simulation. The red dot indicates the location of the monitor point.}
\label{fig:ablation_domain}
\end{figure}

\begin{figure}[htbp]
\centering
\begin{subfigure}[t]{.5\textwidth}%
\centering\captionsetup{width=.8\linewidth}%
\includegraphics[width=\linewidth]{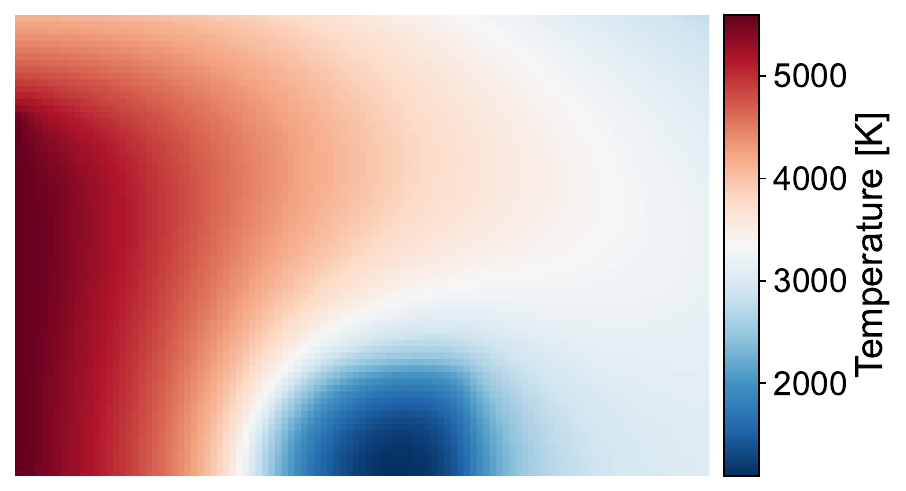}%
\caption{Time-averaged temperature field}
\label{fig:ab_Tfield}
\end{subfigure}
~
\begin{subfigure}[t]{.5\textwidth}%
    \centering\captionsetup{width=.8\linewidth}%
\includegraphics[width=\linewidth]{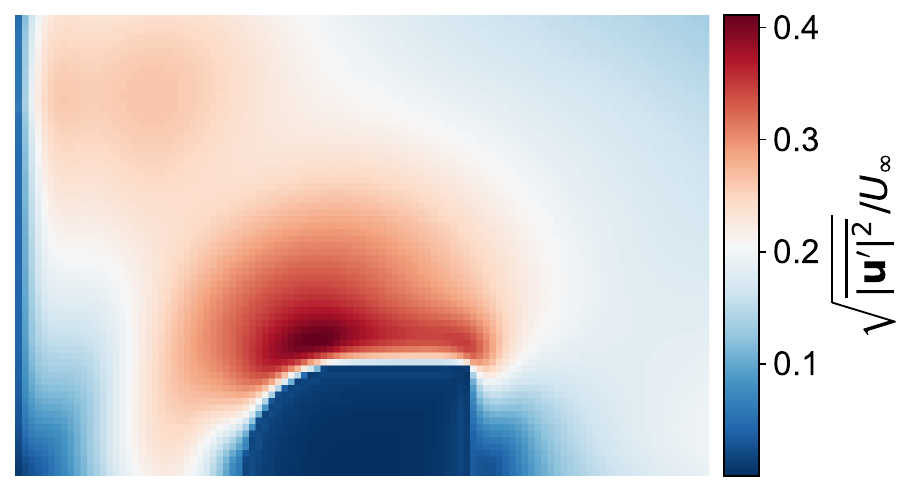}%
\caption{Time-averaged velocity fluctuations field}
\label{fig:ab_vrms}
\end{subfigure}

\caption{Temperature and flow statistics in the physics-based simulations of the ablation study.}
\label{fig:ab_T_vrms_sim}
\end{figure}

\subsubsection{Physics-based model performance \label{subsec:physics-based model}}

In this experiment, a hemispherical sample is exposed to subsonic, high-enthalpy flow generated by an inductively coupled plasma torch in a nitrogen or air environment based on the work of Helber et al.~\cite{helber2015microstructure}.
Since the boundary conditions at the wind tunnel entrance were not provided in the original study, the simulation presented here uses predictions from a separate model \cite{helber2015microstructure} at a certain distance from the sample, as the boundary conditions for the present calculations; therefore, an exact reproduction of the experiment is not possible.
Physical experiments were conducted over a range of inlet pressures and temperatures, producing different inlet velocities.
The results presented here correspond to a nitrogen environment at \SI{1.5}{\kilo\pascal} and a Mach number of $\sim$0.1.

In the setup used here (shown in Figure~\ref{fig:ablation_domain}), flow enters the cylindrical tunnel with a temperature of \SI{5600}{\kelvin}, Reynolds number of 5.2 and composition of $Y_{\ce{N2}}=0.388$ and $Y_{\ce{N}}=0.612$. Reynolds number is defined as $\text{Re}=\frac{\rho U_{\infty}(2R)}{\mu}$ where $U_{\infty}$ is the inlet velocity.
The maximum Courant number is set to 1 with an average of 0.6 to handle the initial unsteadiness, resulting in adjustable time steps sizes on the order of \num{e-6} seconds.
Constant temperature (\SI{350}{\kelvin}) and no-slip boundary conditions are imposed at the tunnel walls and axial symmetry conditions at the centerline.
A constant pressure boundary condition is imposed at the exit.
The domain is discretized using a structured rectangular grid of $270 \times 130$, refined near the object and other boundaries. 
The grid spacing in the close vicinity of the sample is $0.05 R$, with $R=$ \SI{0.0025}{\meter}. 
The sample is made of non-pyrolyzing porous carbon-bonded carbon fiber with an initial density and porosity of \SI{180}{\kilo\gram\per\meter^3} and 0.9, respectively.
The dependence of permeability on porosity is modeled using the linear relationship proposed by Martin~\cite{martin2013volume}.
Thermal and transport properties of the porous material are estimated based on the data and correlation provided by Lachaud et al.~\cite{lachaud2017generic} and Mori~\cite{mori2022laser}.
The nitridation chemistry is approximated by a single-step heterogeneous reaction \cite{helber2017determination}: \ce{C_s + N -> CN}.
Nitrogen recombination/dissociation is also considered using the model by Park et al.~\cite{park2001chemical}; however, within the temperature range considered here, the rate of this reaction remains negligible.

\begin{figure}[htbp]
\centering
\begin{subfigure}[t]{.45\textwidth}%
\centering\captionsetup{width=.8\linewidth}%
\includegraphics[width=\linewidth]{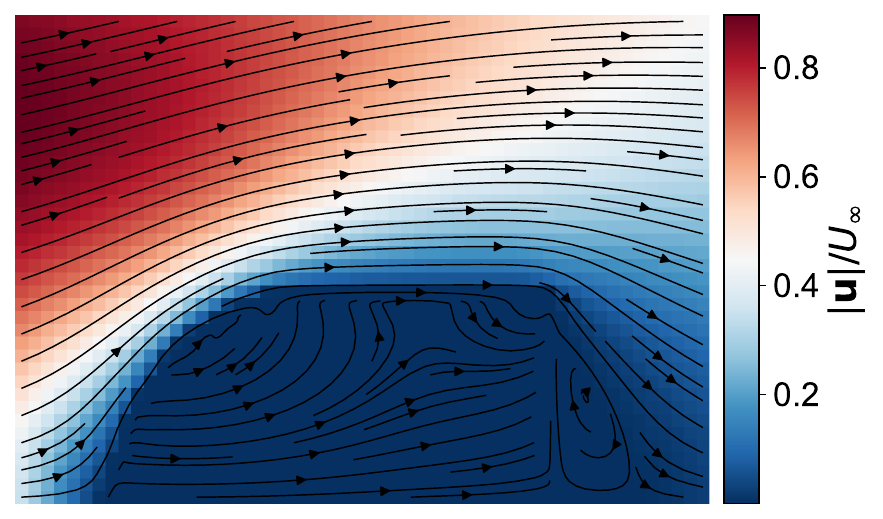}%
\caption{$t=0.57$ \si{s}}
\label{fig:instability_t0.57}
\end{subfigure}
~
\begin{subfigure}[t]{.45\textwidth}%
    \centering\captionsetup{width=.8\linewidth}%
\includegraphics[width=\linewidth]{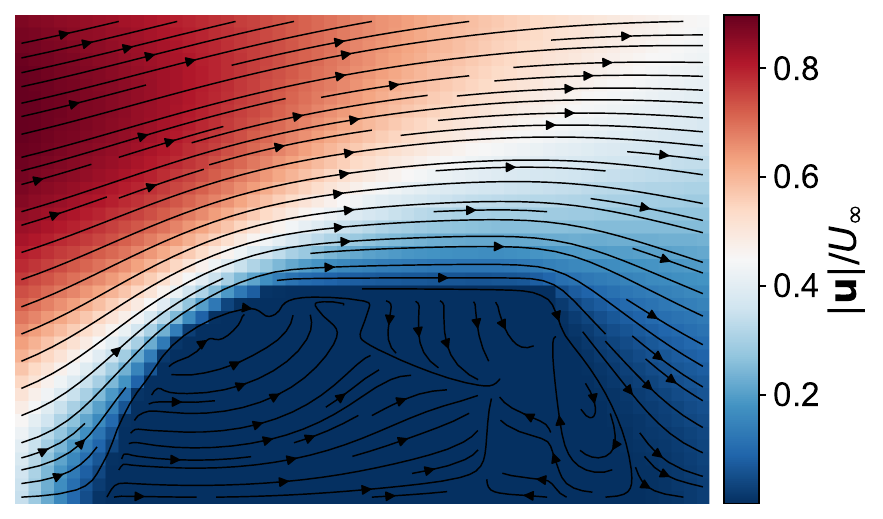}%
\caption{$t=0.589$ \si{s}}
\label{fig:instability_t0.589}
\end{subfigure}

\begin{subfigure}[t]{.45\textwidth}%
\centering\captionsetup{width=.8\linewidth}%
\includegraphics[width=\linewidth]{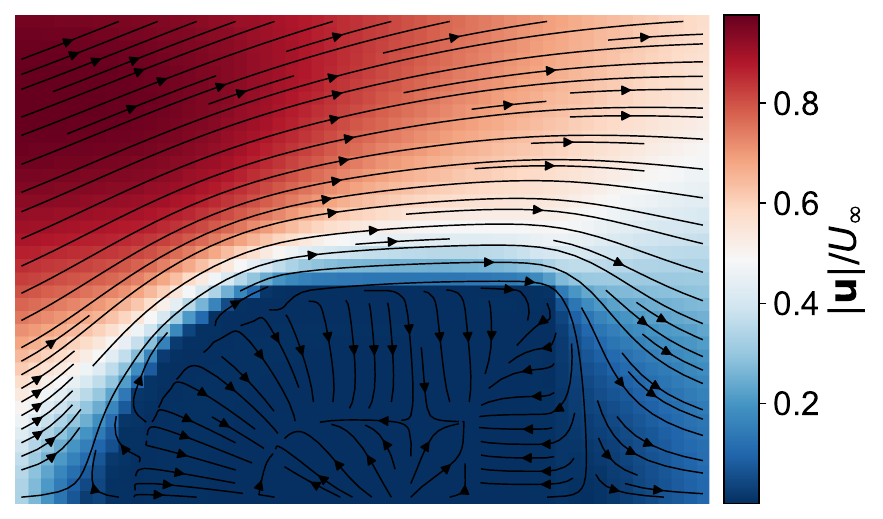}%
\caption{$t=0.597$ \si{s}}
\label{fig:instability_t0.597}
\end{subfigure}
~
\begin{subfigure}[t]{.45\textwidth}%
    \centering\captionsetup{width=.8\linewidth}%
\includegraphics[width=\linewidth]{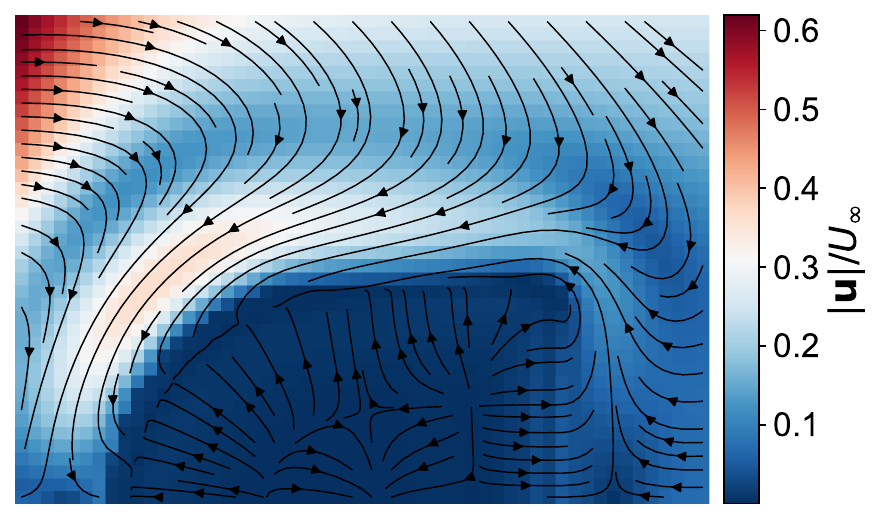}%
\caption{$t=0.5975$ \si{s}}
\label{fig:instability_t0.5975}
\end{subfigure}

\caption{Evolution of flow instabilities in the ablation experiment showing the fields of normalized velocity ($|\mathbf{u}| / U_{\infty}$) in the vicinity of the object.}
\label{fig:ab_vfield_sim}
\end{figure}

\begin{figure}
\centering
\begin{subfigure}[t]{.42\textwidth}%
\centering\captionsetup{width=.8\linewidth}%
\includegraphics[width=\linewidth]{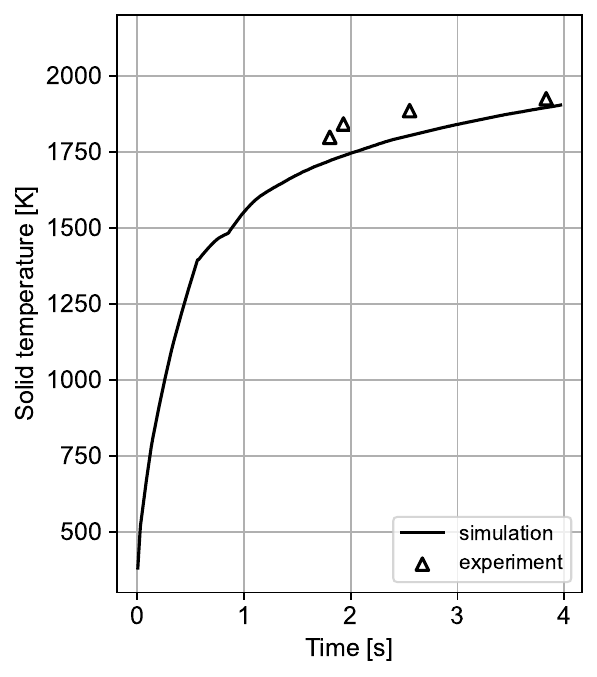}%
\caption{Surface temperature}
\label{fig:ab_Ts}
\end{subfigure}
~
\begin{subfigure}[t]{.41\textwidth}%
    \centering\captionsetup{width=.8\linewidth}%
\includegraphics[width=\linewidth]{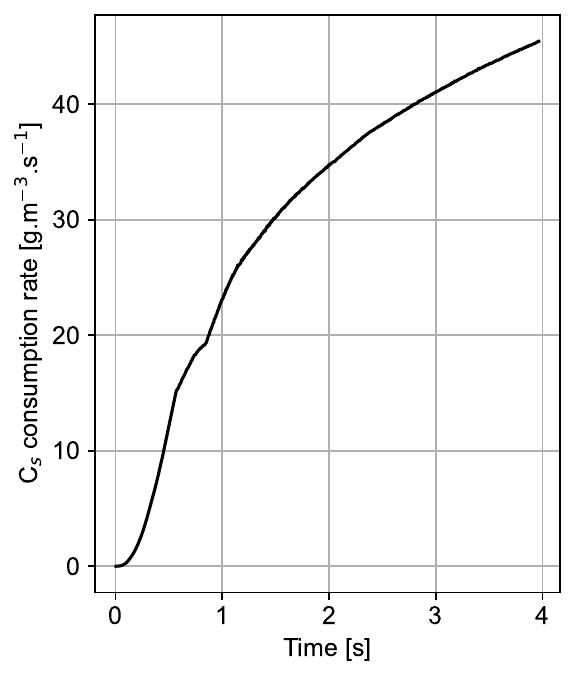}%
\caption{Rate of solid consumption per unit volume of initial sample}
\label{fig:ab_rate}
\end{subfigure}

\caption{Variation of solid temperature and ablation rate at a monitor point on the surface (as specified in Figure~\ref{fig:ablation_domain}), with experimental measurements of solid temperature using a pyrometer taken in a region near the tip of the object from Helber et al.~\cite{helber2015microstructure}.}
\label{fig:ab_Ts_sim}
\end{figure}

Figure~\ref{fig:ab_Tfield} shows the time-averaged gas temperature and the boundary layer that forms in the physics-based simulations.
As the flow impinges on the object, waves of compression and expansion emerge, leading to periodic increases and decreases in pressure and temperature.
These fluctuations originate from the rear of the object and extend toward the front, as Figure~\ref{fig:ab_vrms} shows.
Similar oscillations have been observed in previous studies of flow over bluff bodies \cite{zaman1989natural,almutairi2010intermittent,toppings2025spontaneous}, and have been linked to the quasi-periodic formation and bursting of a laminar separation bubble, which causes the flow to alternate between fully separated and reattached states \cite{toppings2025spontaneous}.
Here, the porous interface coupling and through-flow can add to the complexities of the problem and induce distinct flow instabilities.
Figure~\ref{fig:ab_vfield_sim} shows the evolution of normalized velocity fields, where part of the separated wake is absorbed into the porous medium by suction.
This weakens the base-pressure recovery behind the body and alters the shear balance at the interface. Such conditions can trigger interfacial instabilities, such as Kelvin--Helmholtz, which, as they amplify, feed back into the wake, causing unsteady separation and reverse flow to extend upstream over the body surface.
This phenomenon is closely tied to flow characteristics in the porous region near the interface, influenced by the dynamic pressure difference, and may only occur under certain experimental conditions within this class of problems.

In the solid region, Figure~\ref{fig:ab_Ts_sim} shows the predicted surface temperature and solid consumption rate; the predicted surface temperature falls within 5\% of the measured values.
Once reaction heat release becomes small, the temperature rise slows due to increased heat capacity and radiative losses.
Although the mass loss rates at the early times are not reported in the above reference, the calculated mass loss rate at the end time of the simulation is within 12\% of the overall mass loss rate reported for the physical experiment.

\subsubsection{ROM performance and quantifying uncertainty due to variability in $h_{fs}$}

Similar to the procedure in Section~\ref{subsubsec:non-parametric}, we build a cubic ROM using the following lifted state variables:
\begin{equation*}
    \left[\langle u_x \rangle \;\;\langle u_y \rangle \;\; \langle T \rangle \;\;\langle T_s \rangle \;\; \phi \;\; \frac{1}{\langle \rho \rangle^f} \;\; \langle c_{\textrm{N}} \rangle \;\;\langle c_{\textrm{N}_2}\rangle\;\;\langle c_{\textrm{CN}}\rangle\;\;\langle Y_{\textrm{C}_s}\rangle\right] \;.
\end{equation*}

\begin{figure}[htbp]
\centering
\begin{subfigure}[t]{.3\textwidth}%
    \centering\captionsetup{width=.8\linewidth}%
\includegraphics[width=\linewidth]{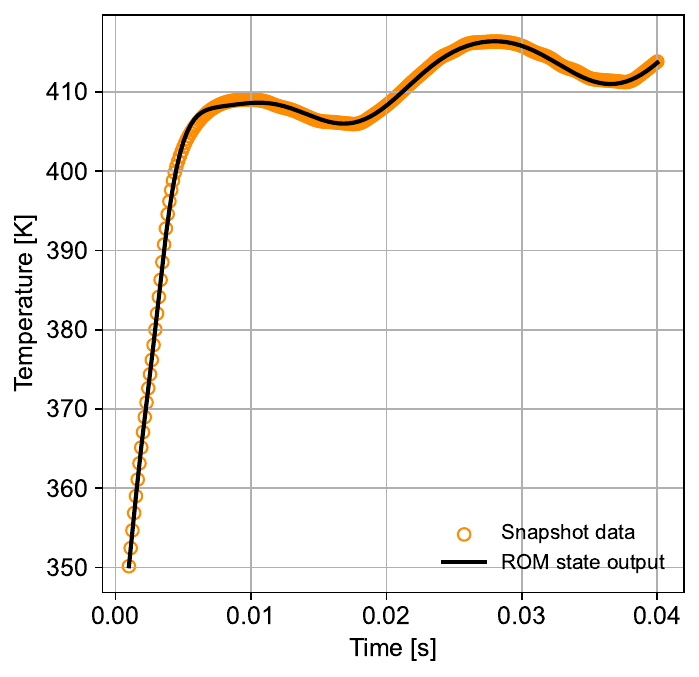}%
\caption{$T$ at ($2.13R,\;0.03R$)}
\label{fig:abTsh011p35}
\end{subfigure}
~
\begin{subfigure}[t]{.3\textwidth}%
\centering\captionsetup{width=.8\linewidth}%
\includegraphics[width=\linewidth]{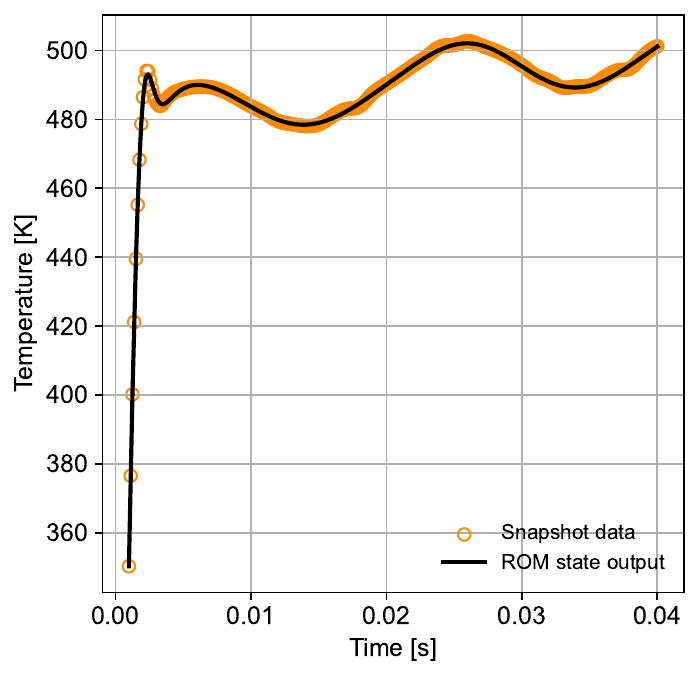}%
\caption{$T$ at ($1.16R,\;0.09R$)}
\label{fig:/abTh011p79}
\end{subfigure}

\begin{subfigure}[t]{.3\textwidth}%
    \centering\captionsetup{width=.8\linewidth}%
\includegraphics[width=\linewidth]{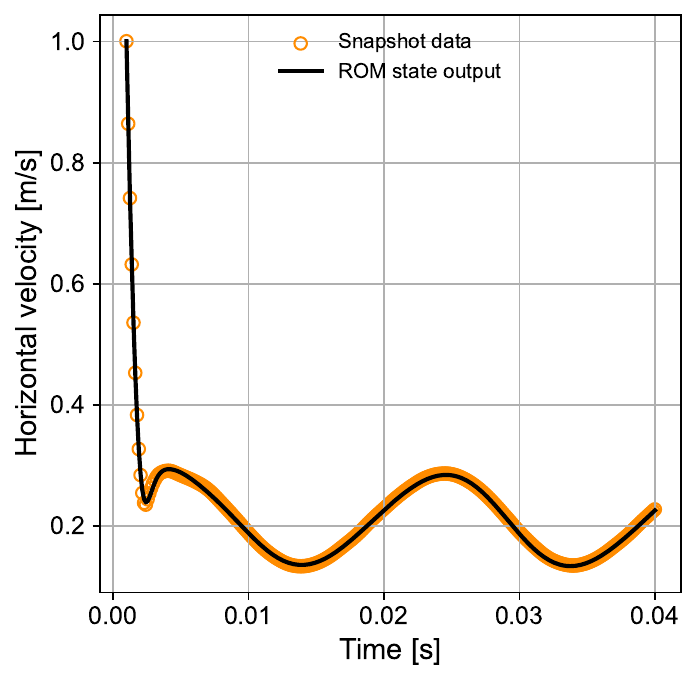}%
\caption{$u_x$ at ($-0.04R,\;0.15R$)}
\label{fig:ab_vx_h011_p120}
\end{subfigure}
~
\begin{subfigure}[t]{.3\textwidth}%
    \centering\captionsetup{width=.8\linewidth}%
\includegraphics[width=\linewidth]{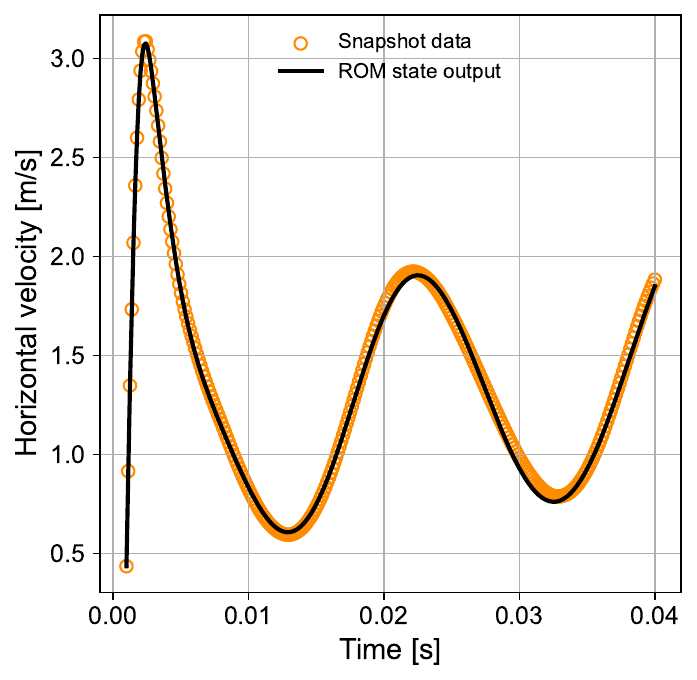}%
\caption{$u_x$ at ($3.25R,\;0.81R$)}
\label{fig:ab_vx_h011_p777}
\end{subfigure}

\caption{Early-time evolution of gas temperature ($T$) and horizontal velocity ($u_x$) at various locations in the domain for ablation case with $h_{fs}=$ \SI{0.11}{\watt\per\square\meter\per\kelvin}, $r=25$, $\lambda_c = 1$, $\lambda_A = 10$, $\lambda_B = \num{1e2}$ and $\lambda_C=\num{1e3}$. The coordinates are measured from an origin at the tip of the object.}
\label{fig:ab_QoI_additional}
\end{figure}

Training data are generated by running full-order simulations over a time span of \SI{0.5}{\second}, with the state saved at 500 uniformly spaced instants ($\Delta t = \SI{1}{\milli\second}$).
The simulations are performed at a pressure of \SI{1.5}{\kilo\pascal}, with the inlet conditions of \SI{5600}{\kelvin} and $\text{Re} = 5.2+1.3\sin (100\pi t)$.
The inlet Reynolds number is modulated sinusoidally at \SI{50}{\hertz} about the nominal value of 5.2, with an amplitude of 25\%, by varying the inlet velocity at fixed inlet temperature, pressure, and composition. 
This forcing is imposed to enrich the transient content of the training data, while keeping the flow in the same subsonic, low-Reynolds-number regime as the constant-inlet case of Section~\ref{subsec:physics-based model}.
The other boundary conditions are the same as those used in the previous section.
A subsection of the simulation domain, defined by a $4.5 R \times 1.5R$ rectangular region surrounding the sample, is used for the analysis, yielding a spatial grid size of $n_{xy}=1400$.
The $k=100$ snapshots spanning $t \in[ \SI{0.3}{\second} ,\; \SI{0.4}{\second}]$ are used to train the ROM, except for the analysis in Figure~\ref{fig:ab_QoI_additional}, which is trained on the $k=40$ snapshots spanning $t \in [0, \SI{0.04}{\second}]$.

\begin{figure}[htbp]
\centering
\begin{subfigure}[t]{.3\textwidth}%
    \centering\captionsetup{width=.8\linewidth}%
\includegraphics[width=\linewidth]{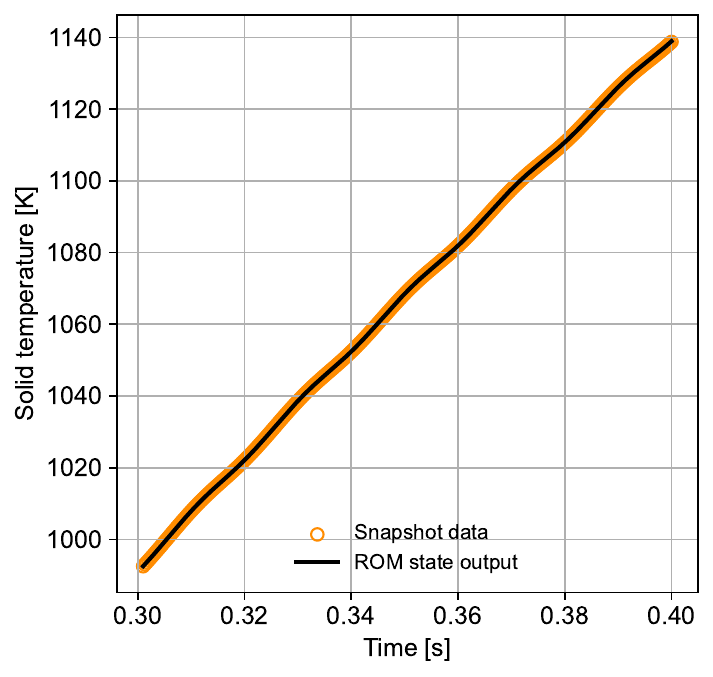}%
\caption{Solid temperature ($T^s$)}
\label{fig:abTsh011}
\end{subfigure}
~
\begin{subfigure}[t]{.3\textwidth}%
\centering\captionsetup{width=.8\linewidth}%
\includegraphics[width=\linewidth]{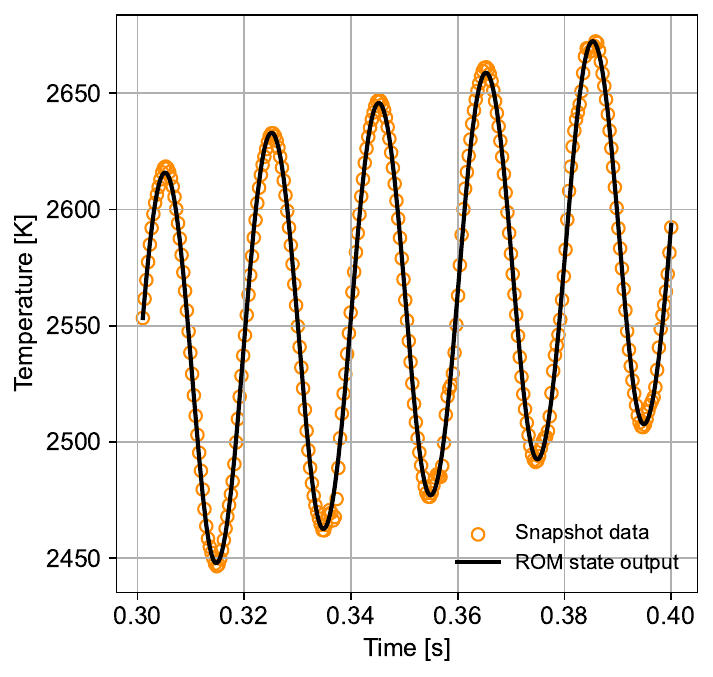}%
\caption{Gas temperature ($T$)}
\label{fig:/abTh011}
\end{subfigure}
~
\begin{subfigure}[t]{.3\textwidth}%
    \centering\captionsetup{width=.8\linewidth}%
\includegraphics[width=\linewidth]{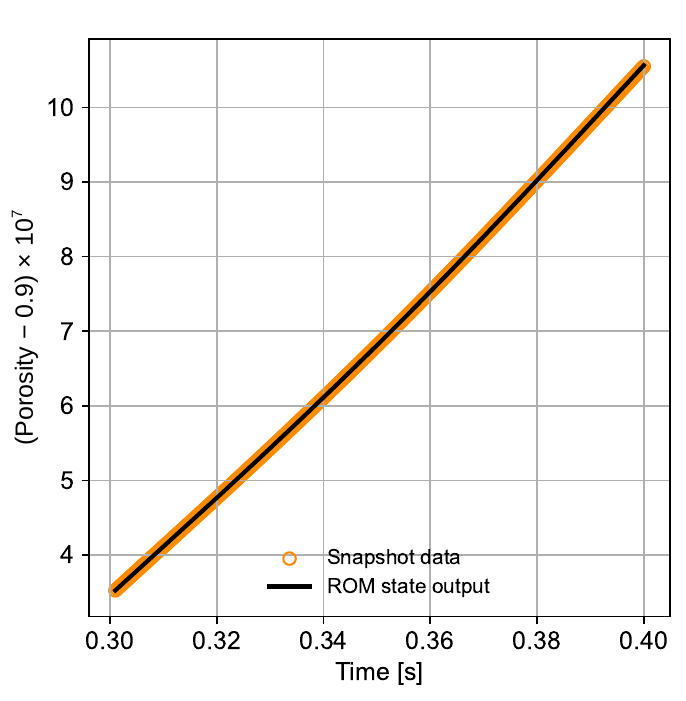}%
\caption{Porosity ($\phi$)}
\label{fig:ab_phi_h011}
\end{subfigure}

\caption{Evolution of quantities of interest at the tip of the object for $h_{fs}=$ \SI{0.11}{\watt\per\square\meter\per\kelvin} with $r=25$, $\lambda_c = 1$, $\lambda_A = 10$, $\lambda_B = \num{1e2}$ and $\lambda_C=\num{1e3}$.
The initial porosity is 0.9; over the training window the porosity increases by roughly \num{7e-7}.}
\label{fig:ab_QoI}
\end{figure}

\begin{figure}[htbp]
\centering
\includegraphics[width=.7\textwidth]{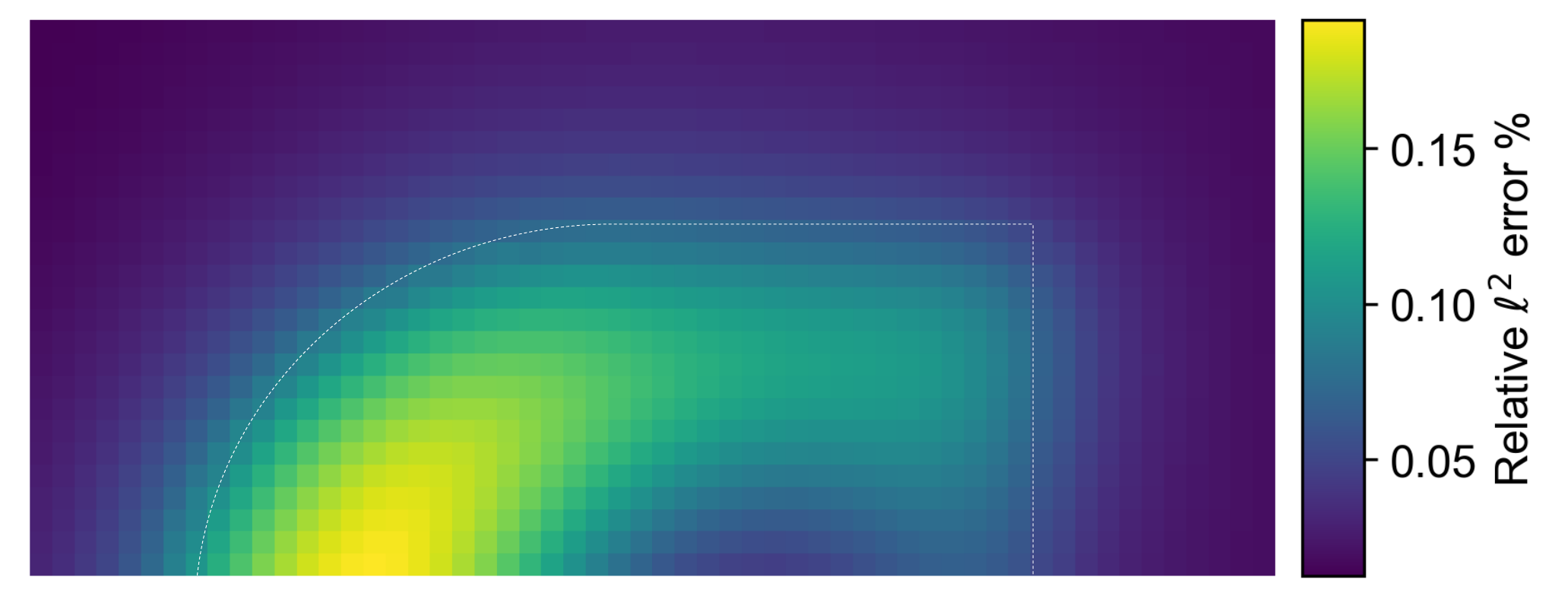}
\caption{Time-averaged spatial distribution of ROM error for  $h_{fs}=$ \SI{0.11}{\watt\per\square\meter\per\kelvin}, $r=25$, $\lambda_c = 1$, $\lambda_A = 10$, $\lambda_B = \num{1e2}$ and $\lambda_C=\num{1e3}$. The dashed line indicates the boundary of the object.}
\label{fig:ablation_error}
\end{figure}

Figure~\ref{fig:ab_QoI} compares the quantities of interest, including solid temperature, porosity, and gas temperature, predicted by the non-parametric ROM and the FOM at the tip of the object.
Figure~\ref{fig:ab_QoI_additional} shows the early-time evolution of gas temperature and horizontal velocity at various locations in the domain.
The comparison shows that the ROM predictions agree well with the training data.
Furthermore, Figure~\ref{fig:ablation_error} presents the time-averaged spatial distribution of overall error (as defined in Section~\ref{subsubsec:non-parametric}), showing that the error remains below 0.2\% over the domain.
The average wall-clock time to integrate the full-order model is approximately \num{1066} seconds.
The parallelized task was executed using 10 CPU cores and 16 GB of memory on a system equipped with 20-core Intel Xeon E5 processors and 128 GB of RAM.
In contrast, the ROMs of size $r = 25$ integrate in $\sim$2.8 seconds, yielding a computational speedup of about $380 \times$.
To our knowledge, this represents the first application of operator inference to a coupled reacting flow--porous media system involving heterogeneous surface chemistry, interfacial heat transfer, and solid decomposition.

To quantify changes due to the uncertainty in solid heat transfer coefficient, we construct a parametric model with the same structure as in Section~\ref{subsec:parametric_model}.
We generate data at each of the 12 parameter realizations $\bm{\eta} = \left(h_{fs}\right)$ for 
\begin{equation*}
    h_{fs} \in \{0.07,\;0.08,\;0.085,\;0.09,\;\dots\;,0.13\} \si{\watt\per\square\meter\per\kelvin} \;,
\end{equation*}
and construct the model by combining five sub-models:
\begin{align*}
    f_{\mathcal{P}_1}(\textbf{q};h_{fs}),\; \mathcal{P}_1 &= \{h_{fs}=0.07,\; h_{fs}=0.09\}: \qquad h_{fs} \in [0.07,\;0.09] \;, \\
    f_{\mathcal{P}_2}(\textbf{q};h_{fs}),\; \mathcal{P}_2 &= \{h_{fs}=0.09,\; h_{fs}=0.10\}: \qquad h_{fs} \in [0.09,\;0.10] \;,\\
    f_{\mathcal{P}_3}(\textbf{q};h_{fs}),\; \mathcal{P}_3 &= \{h_{fs}=0.10,\; h_{fs}=0.11\}: \qquad h_{fs} \in [0.10,\;0.11] \;,\\
    f_{\mathcal{P}_4}(\textbf{q};h_{fs}),\; \mathcal{P}_4 &= \{h_{fs}=0.11,\; h_{fs}=0.12\}: \qquad h_{fs} \in [0.11,\;0.12] \;,\\
    f_{\mathcal{P}_5}(\textbf{q};h_{fs}),\; \mathcal{P}_5 &= \{h_{fs}=0.12,\; h_{fs}=0.13\}: \qquad h_{fs} \in [0.12,\;0.13] \;.
\end{align*}
For a Monte Carlo-like analysis, we draw from a normal distribution for the uncertain parameter:
\begin{equation*}
    h_{fs} \sim \mathcal{N}\!\left(\frac{0.07 + 0.13}{2},\; \left[\frac{0.13 - 0.07}{6} \right]^2 \right) \;.
\end{equation*}
As in the previous case, samples falling outside $[0.07,\;0.13]$ are clipped to the nearest endpoint.

\begin{figure}[htbp]
\centering
\begin{subfigure}[t]{.4\textwidth}%
\centering\captionsetup{width=.8\linewidth}%
\includegraphics[width=\linewidth]{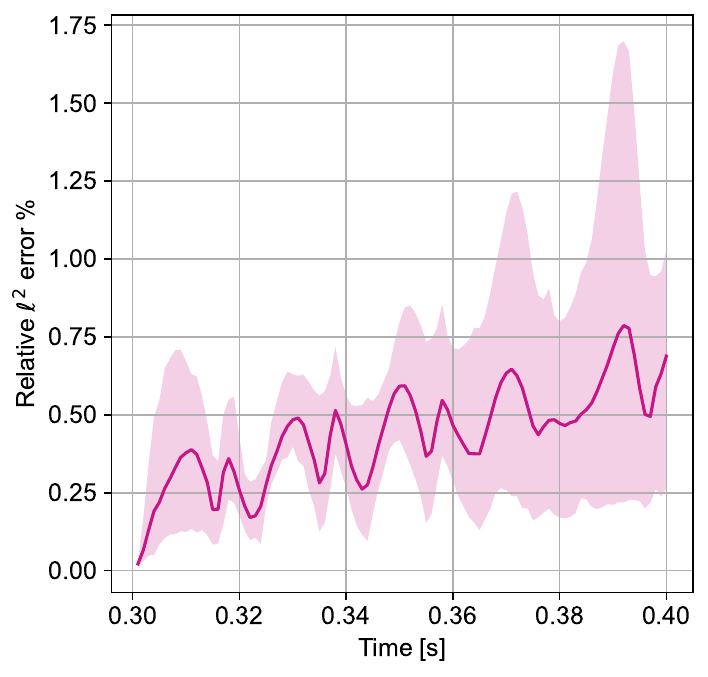}%
\caption{Relative error of parametric ROM averaged over the test data}
\label{fig:ab_errorUQ}
\end{subfigure}
~
\begin{subfigure}[t]{.4\textwidth}%
    \centering\captionsetup{width=.8\linewidth}%
\includegraphics[width=\linewidth]{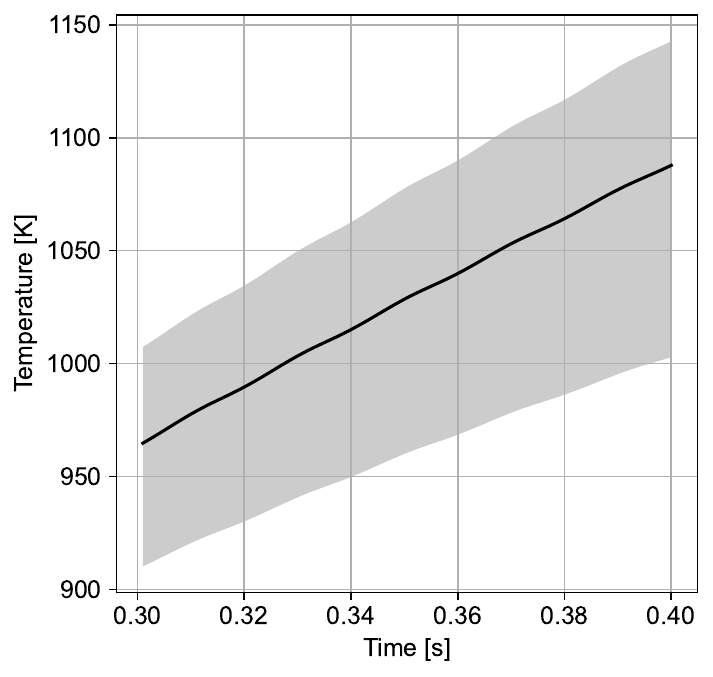}%
\caption{Solid temperature}
\label{fig:ab_TsUQ}
\end{subfigure}

\begin{subfigure}[t]{.4\textwidth}%
\centering\captionsetup{width=.8\linewidth}%
\includegraphics[width=\linewidth]{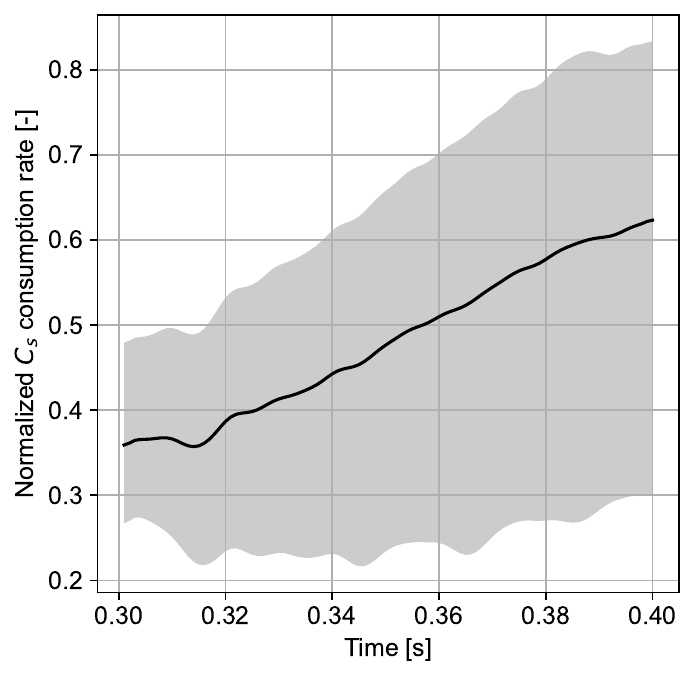}%
\caption{Solid consumption rate}
\label{fig:ab_rateUQ}
\end{subfigure}
~
\begin{subfigure}[t]{.4\textwidth}%
    \centering\captionsetup{width=.8\linewidth}%
\includegraphics[width=\linewidth]{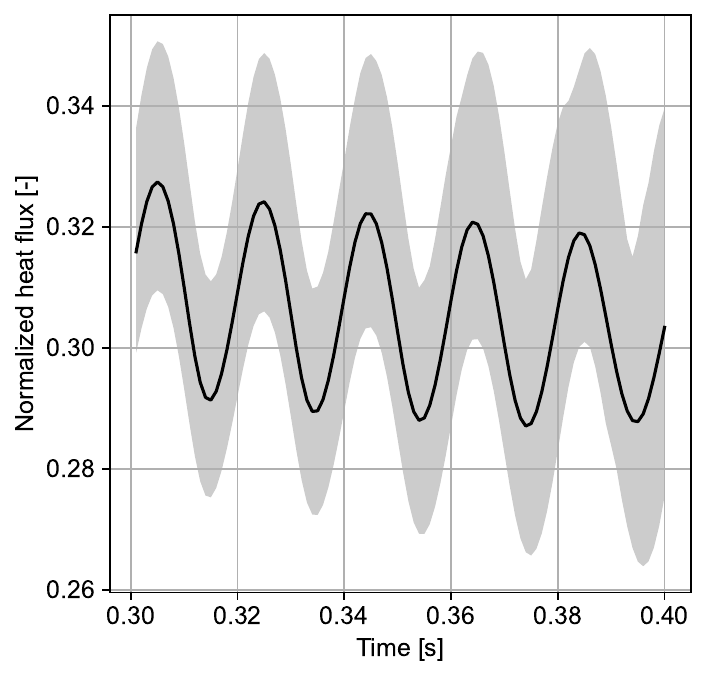}%
\caption{Convective heat flux }
\label{fig:ab_fluxUQ}
\end{subfigure}

\caption{Evolution of the relative overall error averaged over the test data (a) and uncertain quantities of interest at the monitor point on the interface (b--d), with $r=25$, $\lambda = 8\times 10^3$, $k=100$, and $N=150$ for the ablation case. 
The consumption rate and flux are normalized by their maximum value. 
The solid line indicates the mean and the shaded area the standard deviation.}
\label{fig:ab_UQ}
\end{figure}

Figure~\ref{fig:ab_errorUQ} shows the resulting model error evaluated on the test data, which comprises six of the aforementioned parameter realizations, excluding those in $\mathcal{P}_1 \cup \;\dots\;\cup\mathcal{P}_5$.
The propagation of uncertainty through the system is shown in Figure~\ref{fig:ab_UQ}.
Under the conditions of this problem, a variation of approximately 30\% in $h_{fs}$ results in changes of up to $\sim$13\% in the sample surface temperature, $\sim$85\% in solid consumption rate, and up to $\sim$17\% in surface heat flux.
The solid consumption rate directly impacts the heat shield design and, based on the above results, is more sensitive to changes in model parameters than the other variables.

\section{Conclusions}
This work demonstrates that the complex nonlinear system of PDEs---describing coupled processes in a porous solid and the surrounding reacting fluid---can be reduced to a lower-dimensional polynomial form with operators learned from full-order model simulation data.
We also showed that the physics-based model using the single-domain approach is a promising tool for capturing the dynamic interactions at the interface during ablation under high-enthalpy flow conditions.
While the cubic ROM form is an approximation for the reactive source terms, the numerical results indicate that the trained cubic ROM is capable of accurately predicting quantities of interest and preserving species conservation.
However, similar to other data-driven methods, the accuracy of the inferred ROMs highly depends on the training dataset, and it is unrealistic to expect a data-driven ROM to replicate dynamic behaviors that significantly deviate from those seen in the training data.
The lower-dimensional formulation for this problem reduces the system's degrees of freedom by approximately a factor of 560, resulting in a computational speedup of about 380 times.
Beyond demonstrating applicability, this study revealed that parametric ROM quality for this problem class is sensitive to the number and placement of training parameter samples, with fewer well-chosen samples outperforming denser grids---a finding that may inform parameter sampling strategies for operator inference in other multi-physics applications.
The learned parametric ROM successfully captures variations in quantities of interest that come with changes in the permeability and heat transfer coefficient.
The analysis on the effect of the number and location of parameter samples revealed that models built using two-sample models outperform the three-sample models for this problem.
Since the framework proposed here relies on an affine structure in the parametric dependence, future work should focus on extending it to non-affine parametric problems or to cases where the parametric structure is not known. 

While the uncertainty quantification studies presented here consider a single uncertain parameter in each case, the affine parametric structure of the governing equations means that extending the framework to simultaneously quantify uncertainty in multiple parameters---such as permeability and heat transfer coefficient jointly---would require only a modest increase in the number of training simulations.
Exploring such higher-dimensional uncertainty quantification problems is a natural direction for future work.
Additionally, although the parameter domain partitioning adopted here proved effective, determining the optimal number and location of parameter samples to achieve a desired accuracy level remains a challenge for parametric ROMs of this type.

\section*{Acknowledgments}

This project was supported in part by a contract from the Strategic Environmental Research and Development Program (SERDP), Project No. RC19-1092.

\section*{Availability of material}
The software implementations used here are available openly via GitHub under the GNU General Public License version 3 (GPL-3.0), and the version used here was archived on Zenodo~\cite{software, rom-software}. 
Output data and figures are also available openly under a CC-BY license~\cite{repropack}.

\bibliography{ref1}
\bibliographystyle{elsarticle-num}

\end{document}